\documentclass[12pt]{iopart}
\usepackage{graphicx}

\usepackage{soul}
\usepackage{xcolor}

\begin{document}

\title[Design of plasma shutters for improved heavy ion acceleration by ultra-intense laser pulses ]{Design of plasma shutters for improved heavy ion acceleration by ultra-intense laser pulses}

\author{M. Matys$^{1,2}$, S.~V.~Bulanov$^{1,3}$,  M.~Kucharik$^{2}$, M.~Jirka$^{1,2}$, J.~Nikl$^{1,2}$,   M.~Kecova$^1$, J. Proska$^{2}$,  J.~Psikal$^{1,2}$, G.~Korn$^1$  and O.~Klimo$^{1,2}$}

\address{$^{1}$ ELI  Beamlines  Centre,  Institute of  Physics,  Czech  Academy  of  Sciences,  Za  Radnici  835,  25241  Dolni Brezany, Czech~Republic}
\address{$^{2}$ Faculty of Nuclear Sciences and Physical Engineering, Czech Technical University in Prague, Brehova 7, Prague,~115~19,~Czech~Republic}
\address{$^{3}$ Kansai Photon Science Institute, National
	Institutes for Quantum Science and Technology, 8-1-7~Umemidai, Kizugawa-shi, Kyoto 619-0215, Japan}
\ead{Martin.Matys@eli-beams.eu}
\vspace{10pt}
\begin{indented}
\item[]\today
\end{indented}

\begin{abstract}
 In this work, we investigate the application of the plasma shutters for heavy ion acceleration driven by a high-intensity laser pulse. We use particle-in-cell (PIC) and hydrodynamic simulations. The laser pulse, transmitted through the opaque shutter, gains a steep-rising front and its peak intensity is locally increased at the cost of losing part of its energy. These effects have a direct influence on subsequent ion acceleration from the ultrathin target behind the shutter. In our 3D simulations of silicon nitride plasma shutter and a silver target, the maximal energy of high-$ Z $ ions increases significantly when the shutter is included for both linearly and circularly polarized laser pulses. Moreover, application of the plasma shutter  for linearly polarized pulse results in focusing of ions towards the laser axis in the plane perpendicular to the laser polarization. The generated high energy ion beam has significantly lower divergence compared to the broad ion cloud, generated without the shutter. The effects of prepulses are also investigated assuming a double plasma shutter. The first shutter can withstand the assumed sub-ns prepulse (treatment of ns and ps prepulses by other techniques is assumed) and the pulse shaping occurs via interaction with the second shutter. On the basis of our theoretical findings, we formulated an approach towards designing a double plasma shutter  for high-intensity and high-power laser pulses and built a prototype.
\end{abstract}

%\maketitle

\section{Introduction} \label{sec:Intro}
Laser driven ion acceleration is one of the most promising and widely studied features of ultra-intense laser matter interaction, finding applications in various areas \cite{BULANOV2002_onco,Bulanov_2014_UspHadr,Daido_2012_review,Macchi2013_review,Passoni_review_addvanced_2019}. Especially heavy ions are  useful in material science as a radioisotope source and a stage to explore exotic nuclei \cite{Nishiuchi2015,Nishiuchi2016}, in nuclear science involving heavy ion collisions, e.g., for hot and dense matter research  \cite{NuclearRewiew_BRAUNMUNZINGER2016} and schemes like fission-fusion nuclear reaction \cite{Habs2010_FisFus,Lindner2022_Gold}. The heavy ions driven by intense laser can also be used as an injector into conventional accelerators for further research \cite{Korzhimanov2012}. These promising applications drive the goal for improving the laser-accelerated ion beam energy and quality. Introduction of PW-class lasers set up a new records in ion acceleration, e.g. energy of 58 MeV was reached for protons in 2000 \cite{Snavelly2000}, employing the Target Normal Sheath Acceleration (TNSA) mechanism \cite{TNSA_Wilks2001}. The straightforward way to boost the maximal ion energy is to increase the laser pulse intensity. In this way, another mechanisms may get involved especially in combination with the use of ultrathin or low-density targets. Namely, Radiation Pressure Acceleration (RPA) \cite{Esirkepov2004} presents a promising mechanism. Its experimental indications have already been observed  \cite{Henig_Tajima_2009,Kar_2008,Scullion2017,Kim2016_rpa_experiment}. The trend of mechanism blending in high-intensity interaction was demonstrated in 2018 by the experiment involving a hybrid RPA-TNSA regime induced by the onset of the relativistic transparency, setting a new record in proton acceleration of 94 MeV \cite{Higginson2018_100MeV}. The interplay between different ion acceleration mechanisms depends on the target and laser parameters \cite{Bulanov_SS_2016_attain}. For example, the increase of laser intensity for relatively low-density solid targets, like cryogenic hydrogen \cite{Margarone2016}, can result in the shift of the origin of the ions accelerated to the highest energies from the target rear side towards its interior, as demonstrated in Ref. \cite{Psikal2018}. It was also demonstrated in experiment, that energies per nucleon of the bulk carbon ions can reach significantly higher values than the energies of contaminant protons \cite{McIlvenny_2021}. For heavy ions the RPA regime is still-to-be-shown experimentally. Nevertheless, the highest energies are gradually being achieved by using ultrathin targets, e.g., the recent record for gold ion acceleration (exceeding 7 MeV/nucleon) was achieved with lowering the target thickness down to 25 nm \cite{Lindner2022_Gold}. For silver ions, energy exceeding 20 MeV/nucleon was achieved lowering the target thickness down to 50 nm \cite{Nishiuchi2020}. 

Aside the increase of the laser pulse intensity and modification of the target thickness, the maximal energy and quality of laser-accelerated ions can be enhanced by using structured targets, e.g., double-layer \cite{Bulanov2002feas,Bulanov2002DL,Esirkepov2002,Matys2020,Wang_2021,Alejo_2022},  with nano-structures at its surface \cite{Margarone_2012_nanopart}, nano-holes \cite{Psikal2016_hollow,Cantono2021_nanoholes} or a specific geometry, like a transverse Gaussian shape \cite{Chen2009}, a dual parabola \cite{Liu2013_parabola_target}, with a few micron-size holes \cite{Hadjisolomou2020_big_holes} and a pizza-cone target \cite{Gaillard2011_pizza}. 

Another approach is the shaping of the incoming laser pulse.  Techniques like double planar plasma mirrors \cite{Levy2007_plasma_mirror} and others are currently commonly used especially for the improvement of the laser pulse contrast. 
In addition to these techniques, the laser pulse can be also shaped  via its nonlinear evolution as it propagates / burns through an underdense plasma \cite{Bulanov1992Edge,Bulanov1993,Decker1996}, a near critical density plasma \cite{Wang2011_lens,Bin2015_nc_plasma} or an overdense plasma \cite{Vshivkov1998,Reed2009,Palaniyappan2012,Wei2017,Jirka2021_nas}.  In addition to the prepulse treatment, the nonlinear pulse evolution can also result in steepening of the pulse front and (local) intensity increase. The positive effect of these phenomena on ion acceleration in the case of near critical density plasma was demonstrated by experiments and simulations in Ref. \cite{Bin2015_nc_plasma} using a carbon-nanotube foam attached to a solid diamond-like carbon foil, resulting in increase of maximal carbon energy. In this paper we study the high-density (overdense) approach using plasma shutter of thickness relevant to PW class laser systems.

\begin{figure}[ht]
	\begin{center}

		\includegraphics[width=0.65\linewidth]{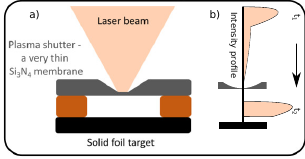}	
		\caption{\label{fig:Scheme} Scheme of the plasma shutter application (a) and the intensity profile of the laser pulse transmitted through the plasma shutter (b). Low intensity parts of the laser pulse are filtered out and the peak intensity increases. }
	\end{center}
\end{figure}

In the context of ion acceleration, the plasma shutter is usually a thin solid foil or membrane which is attached to the front surface of the target with a gap between them (see a scheme in figure \ref{fig:Scheme}(a). 
Therefore, it presents an obstacle which needs to be overcome by the laser pulse and by its accompanying prepulses before the interaction with the main target. Having solid (overdense) density, the plasma shutter is initially opaque for the low-intensity prepulse and beginning of the main pulse. The shutter subsequently becomes relativistically transparent to the rising intensity of the main pulse front. Assuming a thick (semi-infinite) shutter, its electron density $ n_e $ can satisfy relation $ n_e < \gamma n_c $, where $ \gamma $ is the relativistic Lorentz factor (the relation for ultrathin shutter is introduced in the next paragraph). \footnote{ Note that electron density $n_e$ can be locally increased by the radiation pressure. Therefore, a significantly higher initial laser intensity is required for a semi-infinite target to become relativistically transparent than would be given by the initial target density \cite{Cattani_2000,Goloviznin_2000}. On the other hand, the density of thin targets can rapidly decrease by the target expansion and by expelling electrons out of the laser beam axis by the ponderomotive force.} The critical density is defined in CGS units as $n_c =~m_e \omega^2/4\pi e^2 $, where $ \omega $ is laser angular frequency, $ m_e $ and $ e $ are electron mass and charge, respectively. Therefore, the plasma shutter can directly produce a steep-rising front at the beginning of the transmitted laser pulse by filtering out the low intensity parts \cite{Vshivkov1998}. This observation leads to the direct application for ion acceleration as it reduces the target pre-expansion before its interaction with the high-intensity part of the laser pulse. The filtering out of the sub-ns prepulse was demonstrated in experiments and supplementary particle-in-cell (PIC) simulations \cite{Reed2009,Wei2017}. The main target remained overdense in these experiments when the plasma shutter was included, resulting in the increase of ion energy. This prepulse reduction can be especially important for the use of nanostructures / nanoholes in the target, as they diminish for low-contrast lasers otherwise  \cite{Cantono2021_nanoholes}. Moreover, the steepening of the main pulse itself becomes important with the increasing intensity as it significantly affects the laser-electron dynamics in ultrathin targets as is shown in this work. The steep-rising front of the intensity profile can also mitigate the development of transverse short-wavelength instabilities usually ascribed to Rayleigh-Taylor instability \cite{Pegoraro2007,Klimo2008PRST,Robinson2008} or electron-ion coupled instability \cite{Wan2016,Wan2018}. The mitigation of its development was demonstrated in \cite{Matys2020} by numerically steepening of the pulse front. The steep front can also enhance the photon emission from under-dense targets \cite{Jirka2020}. The ultra-thin shutter also provides ideal conditions for the research of the relativistic induced transparency itself \cite{Palaniyappan2012}, leading to the formation of a relativistic plasma aperture \cite{Gonzales2016electrons}. The incident laser light is diffracted at this aperture having effects on electron \cite{Gonzales2016electrons} and ion \cite{Gonzalez2016proton} distributions. The diffraction of the laser pulse on such an aperture and its constructive interference with generated high harmonics can lead to a local intensity amplification \cite{Jirka2021_nas}. Therefore, the use of the plasma shutter (or a series of them as discussed in section \ref{sec:DoublePSh}) can provide three positive effects as is sketched in figure \ref{fig:Scheme}(b): 1)~mitigation of prepulses, 2)~generation of the steep-front laser pulse, 3)~local increase of the peak intensity. The negative effect is the partial lost of the laser pulse energy, depending on the shutter thickness. Note, that the use of a single plasma shutter for prepulse reduction inherently decreases its density, making the later two effects less significant (as discussed below). Therefore, a series of plasma shutters (each focusing on different effect) can provide a way to use the full potential of such interaction.

The interplay between a relativistic ($ a_0>0 $) laser pulse and a thin plasma shutter depends on two main parameters, the dimensionless amplitude of the laser electric field $ a_0 = eE_0/m_e\omega c $ and the areal density of the target $ \epsilon_0 = \lambda l/4\pi l_s^2 = \pi\frac{n_e}{n_{c}}\frac{l}{\lambda} $, where $ E_0 $ is the electric field amplitude, $ c $ is speed of light in vacuum, $ \lambda $ is the wavelength of the incident laser pulse, $ l $ is the thickness of the plasma shutter and $ l_s $ is the skin depth defined as $ l_{s} = c/\omega_{pe}$ for the plasma electron angular frequency $ \omega_{pe} $. The thin plasma shutter becomes relativistically transparent to parts of incident laser pulse fulfilling the condition  $ a_0 \gg \epsilon_0 $ \cite{Vshivkov1998}. Therefore, the increase of the shutter thickness reduces the energy of transmitted laser pulse. On the opposite, the local peak intensity of transmitted laser pulse increases with the shutter thickness \cite{Jirka2021_nas}, saturating around the relativistically corrected skin depth $l_{sc}=\sqrt{\gamma}c/\omega_{pe}$. Therefore, the thickness of the plasma shutter needs to be optimized assuming both of these effects for efficient ion acceleration as they act against each other. More laser pulse energy will be transmitted through the thinner plasma shutter, but the intensity increase will be lower and vice versa. 
 
 The transmitted laser pulse (shaped by the plasma shutter) is subsequently used for ion acceleration from the target placed behind the plasma shutter. The optimal target thickness $ l $ for ion acceleration via RPA mechanism \cite{Esirkepov2004} (for which $ a_0=\epsilon_0 $)  can be expressed as:
  \begin{equation} \label{eq:opt}
  \frac{l}{\lambda}=\frac{a_0}{\pi}\frac{n_c}{n_e}
  \end{equation} 

In this work, we investigate the application of the plasma shutters for heavy-ion acceleration driven by a high-intensity laser pulse using 3D and 2D PIC simulations. Firstly, we assume an idealized situation with a plasma shutter being opaque for the high-intensity part of the laser pulse without any prepulse treatment. The main laser pulse, transmitted through the opaque plasma shutter, gains a steep-rising front and its peak intensity is locally increased at the cost of losing part of its energy, depending on the shutter thickness as was recently demonstrated in \cite{Jirka2021_nas} (e.g. over 50~\% of the main pulse energy was lost for the effect of local pulse intensity increase by the factor of 7 in Ref. \cite{Jirka2021_nas}). Therefore, we investigate a possible application of such shaped pulse for ion acceleration from the ultrathin target located behind the plasma shutter, where the advantages of pulse profile modifications need to overcome the laser pulse energy loss.  Subsequently we investigate a more realistic scenario including interaction with a sub-ns prepulse with intensity relevant to nowadays PW-class laser systems.  For this scenario we propose the use of two (or eventually multiple) shutters for ion acceleration having different purposes. The first shutter will be used for the prepulse treatment in the way already demonstrated in Refs \cite{Reed2009,Wei2017}, becoming transparent to the main pulse and thus allowing high percentage of transmitted pulse energy (up to 99\% in Ref. \cite{Reed2009}). The second shutter, being opaque for the main pulse, can directly shape the main pulse \cite{Jirka2021_nas}. The whole double-shutter scenario is investigated using a combination of 2D hydrodynamic (simulating the prepulse interaction with the first shutter) and PIC simulations.

The parameters like target and shutter thicknesses, length of the gap between them, and the effects of the pulse-front steepness are investigated using 2D PIC simulations for the case of a silicon nitride plasma shutter and a silver target.  The silver target was chosen for demonstration of our heavy-ion scheme assuming a PW-class laser, because its optimal thickness for RPA mechanism (eq. \ref{eq:opt}),  is relatively high compared to materials with higher $ Z $, as this thickness decreases with increasing target density. Therefore, the targets with thickness of the order of the optimal one (lower 10s of nm) are reasonably thick for future experiments (silver targets with thickness around 50 nm have been already used \cite{Nishiuchi2020}.). The relatively high thickness also makes the 3D PIC simulations less computationally demanding, as the target can be still resolved by at least a few cells even for higher cell size.
 
  In the follow-up 3D simulations of our 2D reference case, the maximal energy of silver ions increases by 35\% when the plasma shutter is included (2D simulations with a lower $ Z $ target (aluminium) were also performed, resulting in similar but 
less significant increase.). Moreover, the steep-rising front of the laser pulse leads to formation of high density electron bunches. This structure (which also appears in the generated electric and magnetic fields) focuses ions towards the laser axis in the plane perpendicular to the laser polarization. The generated high-energy ion beam has significantly lower divergence compared to the broad ion cloud, generated without the shutter. In the later case, the structures are pre-expanded with the low intensity part of the laser pulse and the subsequent onset of the transverse instability mentioned above. The effects of prepulses are investigated using a combination of 2D PIC and hydrodynamic simulations assuming a double plasma shutter. The first shutter can filter out the assumed sub-ns prepulse (treatment of ns and ps prepulses by other techniques is assumed). Therefore, the processes of the steep-front generation and the local intensity increase can develop via interaction with the second non-expanded shutter. The increase of the maximal ion energy is demonstrated also in this case. A prototype of this double-shutter is presented.
 
 The paper is organized as follows. The simulation method and
 parameters are described in section \ref{sec:parameters}. Section \ref{sec:Results}, containing results, is divided into five subsections. Firstly,  the modifications of the laser pulse transmitted through the plasma shutter, like its local intensity increase and energy loses, are studied in section \ref{sec:Laser_shaping}. The optimal shutter and target thicknesses  and length of the gap between them for ion acceleration are discussed in section \ref{sec:thickness_tar}. The influence of the generated steep-rising front on ion acceleration is investigated via 2D simulations using approximated pulse with different steepness in section \ref{sec:Steep:front}. Then the whole picture is discussed via 3D simulations comparing the cases of silver target with and without the plasma shutter for linear and circular polarization in section \ref{sec3D}. Lastly, the use of double plasma shutter is discussed using the combination of 2D hydrodynamic and PIC simulations, taking the prepulse into account, in section \ref{sec:DoublePSh}. The prototype of such a shutter is also presented in that section.   
  
\section{Simulation method and parameters} \label{sec:parameters}

To demonstrate the advantages coming from implementation of the plasma shutter on ion acceleration we performed 2D and 3D particle-in-cell simulations using the code EPOCH \cite{Arber2015}. 

The parameters of the 2D simulations are as follows. Linearly p-polarized (electric field lies in the plane of incidence) laser pulse incidents normally on the target.  The radiation wavelength is $ \lambda = 1\ \mathrm{\mu m}$ and the peak intensity is $ I_{\mathrm{max}} = 10^{22}\  \mathrm{W/cm^2}$, thus yielding dimensionless amplitude $ a_0 \approx 0.85\sqrt{I\left[10^{18}\mathrm{W\ cm^2}\right]\lambda^2\left[\mathrm{\mu m}\right]}  \approx 85 $. The critical plasma density is equal to $ n_c \approx 1.115\times10^{21}\  \mathrm{cm^{-3}}$.

 The laser pulse has a Gaussian spatial profile with beam width at the full width at half maximum (FWHM) equal to 3 $\lambda $. The temporal profile has $ \sin^2(t) $ shape in intensity and beam duration is 64 fs. This pulse corresponds to a 30 fs long 1 PW laser pulse with Gaussian envelope. The laser pulse energy can be fully represented by the $ \sin^2(t) $ profile in the PIC simulations, avoiding the numerical cut of the infinite exponential beginning of the Gaussian shape. Therefore, a further analysis of transmitted energy is more reliable. 

The plasma shutter is made of silicon nitride ($ \mathrm{Si_3N_4} $) solid foil. Full ionization of the foil is assumed with electron density $ n_{e} = 835\ n_c $. In the reference case the thickness of the plasma shutter is set to 20 nm. Plasma shutters with thickness between 12 nm and 45 nm were also considered. Thus, the interplay between parameters $ a_0 $ and $ \epsilon_0 $ discussed in section \ref{sec:Intro} is as follows: $ \epsilon_0 = a_0 = 85$  for shutter thickness $ l \approx 32.5\ \mathrm{nm}$, for the reference case of $ l = 20\ \mathrm{nm}$ then $ \epsilon_0 \approx 52.5$ and for the thickness equal to the relativistically corected skin depth $l=l_{sc}=42.7$ nm,  $ \epsilon_0 \approx 112$.

The target which is located behind the plasma shutter corresponds to commercial silver solid foil with thickness of 20 nm. For the discussion of optimal thickness see the section \ref{sec:thickness_tar}. Partial ionization of the target is assumed, which is in agreement with the experiments of similar type of laser with a silver target \cite{Nishiuchi2020}. For simplicity, the target consists of electrons with density $ n_{e} = 2100\ n_c $ and silver ions with charge number $ Z=40 $ and mass number $ A = 108 $. 

Two sets of simulations were performed. Firstly, the simulation was done with the box size of $ 44\ \lambda\times 17\ \lambda$ ($\times 17\ \lambda$ for 3D cases) focusing on laser pulse transmission through the plasma shutter. Plasma shutter is placed in the middle of the simulation box, referred to as position $ x = 0 $. This setup ensures that both the reflected and transmitted part of the laser pulse are fully captured in the simulation for comparison. Subsequently, the simulation box is prolonged to the size of $ 60\ \lambda\times 17\ \lambda$ ($\times 17\ \lambda$ for 3D cases) including ion acceleration from the target. In this case the plasma shutter is placed at the position $ x = 0 $, situated $ 12\ \lambda $ from the
simulation box boundary in the direction of the laser propagation. The transverse size of the plasma shutter and the target is $ 17\ \lambda $, i.e. the target
is reaching the simulation box boundaries at positions
$ y = \pm\ 8.5\ \lambda$ where thermal boundary conditions for particles are applied.

In 2D simulations, the mesh has square cells of the size $ 0.003\ \lambda $. This is shorter than the plasma skin depth $ c/\omega_{pe}\approx 0.0055\  \lambda $, where $ \omega_{pe} $ is electron plasma frequency corresponding to electron density of the shutter (the cell size is also lower than the plasma skin depth of the target $ c/\omega_{pe}\approx 0.0035\  \lambda $, using the electron density of the target). Since the 3rd order b-spline shape of the quasi-particles and current smoothing are used in our simulations, it is ensured that numerical heating is strongly reduced even for the cells larger than the plasma Debye length \cite{Arber2015}. 
The number of electrons is set to correspond to 835 particles per cell inside the plasma shutter and 1050 inside the target, respectively. The numbers of ions correspond to their charge ratios.

In 3D simulations, the mesh has cuboid cells. The size of the cells is set to $ 0.005\ \lambda $ in the laser propagation direction ($ x $- direction) and $ 0.025\ \lambda $ in transverse directions ($ y$- and $ z$- directions). Number of electrons per cell was reduced to 400 inside the plasma shutter and 1000 inside the target. The ratios for number of ions are kept the same as in the 2D case. 

Temperatures of all particles are initialized to 5 keV inside the plasma shutter and to 0.5 keV inside the target, to further reduce numerical heating. The particle solver begins to move the particles just a few time steps before the arrival of the laser pulse front to the plasma shutter. 

In order to include the effect of target pre-expansion done by the prepulse, the particle-in-cell simulations were combined with the output of hydrodynamic simulations performed in the code PALE (Prague Arbitrary Lagrangian-Eulerian) \cite{Liska2011,Kapin2008}, in section \ref{sec:DoublePSh}. This code employs a 2D cylindrical geometry in a moving Lagrangian framework. The robustness of the simulation is improved by performing periodic mesh smoothing followed by a conservative interpolation of the state quantities from the Lagrangian to the smoothed mesh \cite{Kucharik2003}. The simulations involve a heat conductivity numerical model \cite{LiskaEQUADIFF} and heat flux limiter, simple critical-density and wave-based laser absorption models \cite{Velechovsky2015phdthesis}, Spitzer-Harm heat conductivity model \cite{Spitzer1953}  and realistic Quotidian Equation of State (QEOS) \cite{More1988}. This code is commonly used for interpretation of experiments involving nanosecond pulses \cite{Badziak2016,Picciotto2014,Margarone2014}. The simulation uses a 2D computational $r$-$z$ mesh containing $100\times 100$ cells including a $40\,\mathrm{nm}$ thick and $25\,\mu\mathrm{m}$ long foil of solid aluminum at room temperature. The areal density is roughly the same as of the silicon nitride plasma shutter. Another set of simulations contains a corresponding $20\,\mathrm{nm}$ thick silver target. The computational cells are geometrically distributed in both dimensions to achieve high mesh resolution in the absorption region. A constant temporal intensity of $10^{12}\, \mathrm{W}/\mathrm{cm}^2$ and Gaussian spatial laser beam profile with the focus radius $r_f=2\  \mu\mathrm{m}$ are considered.

\section{Results} \label{sec:Results}

\subsection{Laser pulse shaping via plasma shutter} \label{sec:Laser_shaping}
The properties of the laser pulse transmitted through the plasma shutter of certain material depend on the thickness of the plasma shutter and differ with the dimensionality of the simulation. Therefore, the variables like intensity increase, position of the focus and the percentage of transmitted energy are discussed in this section.

\begin{figure}[ht]
	\begin{center}
		\flushleft
		\includegraphics[width=1\linewidth]{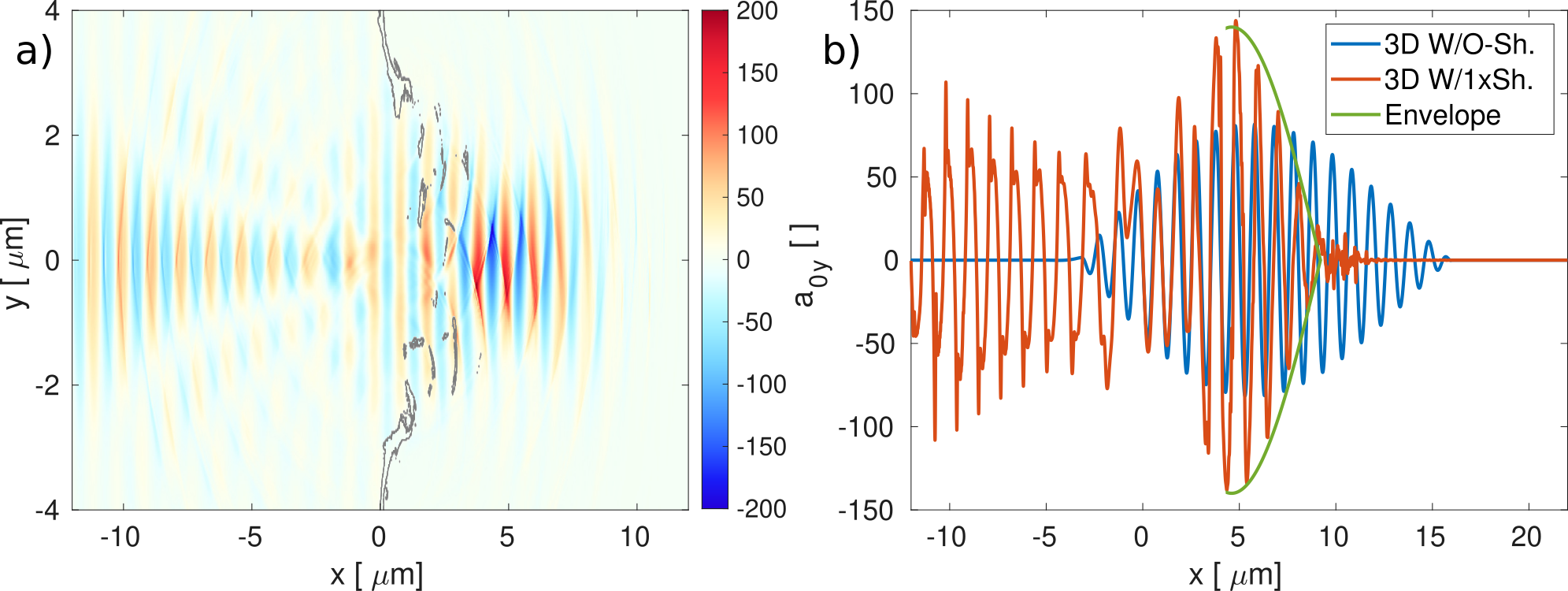}	
		\caption{\label{fig:Sh3D_slices} Increase of the amplitude of the electric field in the $ y $-direction ($ a_{0y} $) after the laser pulse propagates through the plasma shutter in the 3D simulation. (a) 2D slice at $ z=0 $, gray contours depict the areas with electron density $n_e \geq 10\ n_c$. (b) 1D slice at $ y=0 $ and $ z=0 $, comparison of the transmitted laser pulse (W/1xSh) with the original one (W/O-Sh) and the sinusoidal envelope described in the text.}
	\end{center}
\end{figure}

The properties of our reference case, using the 20 nm thick plasma shutter from the 3D simulation, which is then used for ion acceleration, are shown via 2D and 1D slices in figure \ref{fig:Sh3D_slices} at time $ t = 38\ T$ (from the beginning of the simulation). 
The electric field in the $ y$-direction (parallel to the laser polarization) is presented in the form of $ a_{0y}= eE_y/m_e\omega c $. The laser pulse with initial $ a_0 = 85 $ burns through the plasma shutter located at $ x = 0\ \mathrm{\mu m}$. The relativistic aperture is developed as can be seen via contours of density $ n_e\ge 10\ n_c $ in figure \ref{fig:Sh3D_slices}(a). It leads to the diffraction of the transmitting laser pulse and its local constructive interference with generated high harmonics as was described in Ref. \cite{Jirka2021_nas}. The value of $ a_{0y} $ can locally increase from the original $ a_{0y} = 85 $ to over $ a_{0y} = 200 $. The local maxima are located off-axis alternating $ \pm $ in the $ y $-direction, where the interference happens. The central 1D profile (at $ y=0 $ and $ z=0 $) is shown in figure \ref{fig:Sh3D_slices}(b). The transmitted laser pulse acquires the steep-rising front at the beginning. Its approximation using the equation $ y=140*(\sin(\pi*(x)/(9.2))) $ is depicted with the green color. Thus, the rising front is about 5 $T$ shorter compared to the original pulse (depicted by the blue line) with duration of 19.2 T. Moreover, the steepness of the rising front is increased by the rise in the intensity.  Note that the distribution of the laser pulse electric field (and also its energy) differs in space from the case without the shutter (pure Gaussian), being more dense around the central region  as can be seen in  figure 2a.  The integrated energy density of the transmitted laser pulse electric field in the central 1D profile (evaluated forwards from 0) is actually about 42\% higher compared to the simulation without the shutter, even though a significant part of the laser pulse is reflected from the plasma shutter (figure \ref{fig:Sh3D_slices}(b)). On the contrary the total (whole space) transmitted pulse energy is about 50\% lower as discussed below.   
 
\begin{figure}[ht]
	\begin{center}
		\includegraphics[width=\linewidth]{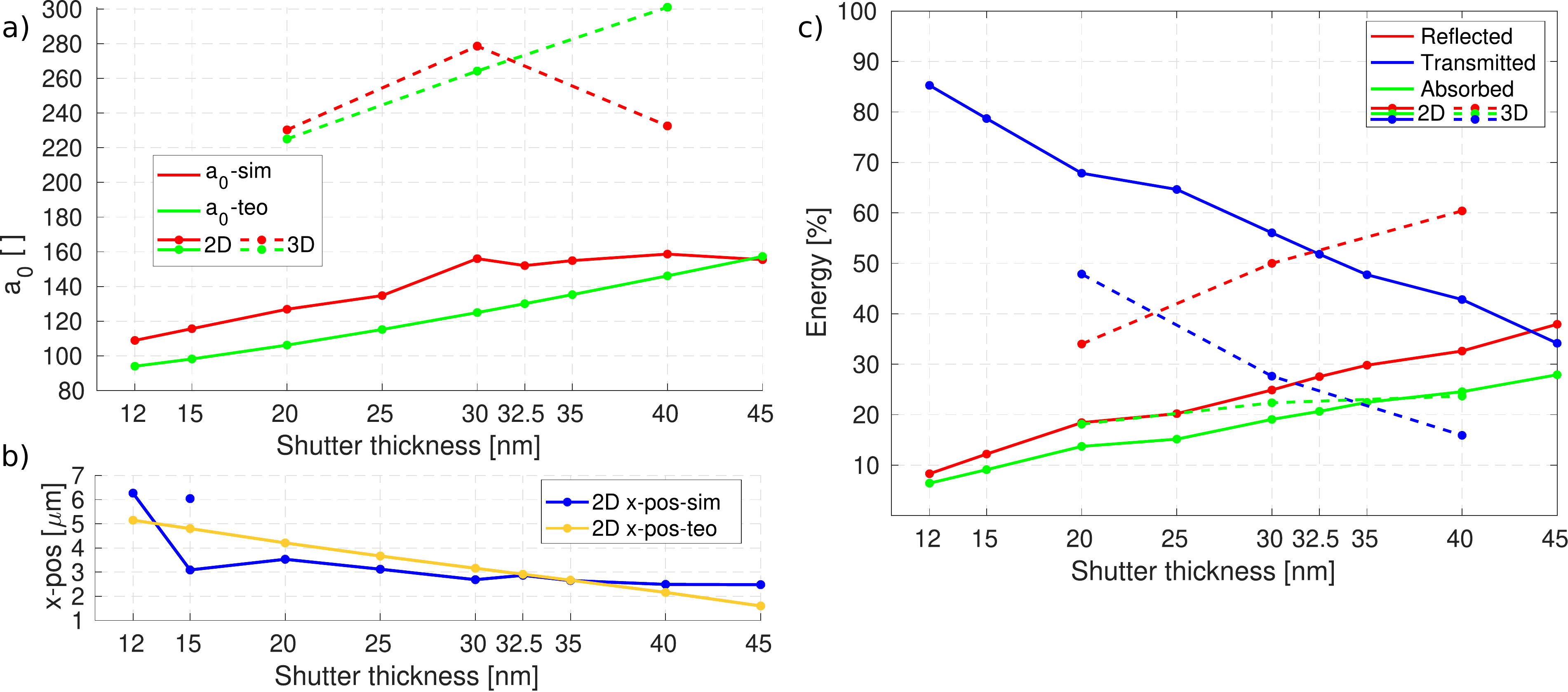}	
		\caption{\label{fig:ParamSh} Comparison with theory \cite{Jirka2021_nas} of achieved cycle averaged maximum of (a)~the~transmitted pulse $ a_0 $, (b)  the $ x $-position of its focus for different shutter thickness. (c) Energy balance of reflected, transmitted and absorbed energy for different shutter thickness.}
	\end{center}
\end{figure}

 The comparison of maximal reached $ a_0 $ and the $ x $-position of its focus from 2D PIC simulations with the theoretical model \cite{Jirka2021_nas} for different shutter thicknesses are shown in the lower part of figure \ref{fig:ParamSh}(a). The values represent the cycle averaged maximum of $ a_0$ obtained from the simulation data.
The intensity is rising with thickness, while the $x$-position of the focus is decreasing. After reaching a local maximum at 30 nm, the maximal $ a_0 $ is gradually saturating, reaching a global maximum at 40 nm and slowly decreasing afterwards. The $x$-position of the focus differs a little after saturation. This observations are in agreement with theory \cite{Jirka2021_nas} for "p" polarization as the electric field amplification should be saturated after the shutter thickness reaches the relativistically corrected skin depth $ l_{sc}$, which is approx. 42.7 nm for our case. The prediction of the focus $ x $-position for ultra-thin targets below 20 nm is challenging as the focus position changes with time and the interference pattern may produce several maxima at different positions. As an example, the second maxima for the case of 15 nm is shown by the blue dot. The predicted focus is roughly in the middle of these two values.

In 3D simulations, shown in the upper part of figure \ref{fig:ParamSh}(a), the maximal $ a_0 $ fits well to the theory and rises with thickness between 20 nm and 30 nm cases. The saturation process is not observed in the simulation of 40 nm shutter as the maximal intensity decreases to value similar to the 20 nm case. The increase of the maximal $ a_0 $ is almost two times higher in the 3D simulations compared to their 2D counterparts, which corresponds to the sum of two processes focusing the beam either in the polarization plane ('p') or out of it ('s') \cite{Jirka2021_nas}. On the contrary, focusing in only one dimension occurs in the 2D cases. The highest value of $ a_0 $ is observed in the case of 30 nm, increasing from $ a_0 = 85 $ to $ a_0 = 280 $. Thus, yielding the intensity increase by the factor of $I_{max}/I_{init} = (a_{0max}/a_{0init})^2 \approx  10.85$. The absence of saturation effect in the case of 40 nm shutter in 3D can be ascribed to the combination of 's' and 'p' polarizations as the 's' part do not provide a saturation effect even in the 2D cases as was shown in \cite{Jirka2021_nas}. Therefore, the optimal thickness for the highest local intensity increase in 3D is in our case around the condition of $ a_0=\epsilon_0 $.

Another important parameter apart the intensity increase is the percentage of transmitted laser pulse energy. The energy balance for different thicknesses of the plasma shutter is investigated in figure \ref{fig:ParamSh}(c). The simulation box is set to be able to contain the whole incident and transmitted laser pulse.

In 2D simulations (solid lines) the transmitted laser pulse energy is decreasing with thickness but stays above 50\% for thicknesses where $ a_0>=\epsilon_0 $ (thickness $ <= 32.5 $ nm), as the plasma shutter becomes partially (relativistic) transparent before the intensity maximum reaches the plasma shutter \cite{Vshivkov1998}. In the 3D cases (dashed lines) the reflection and partially absorption are stronger than in the 2D cases, where the spherical apertures are not limited in the third dimension. About 50\% of the laser pulse energy is transmitted in the case of 20 nm, decreasing to 28\% for the 30 nm case and 16\% for the 40 nm case. As the laser pulse energy loses (absorption and reflection) are significant with the increasing thickness in 3D, their influence needs to be taken into account for choosing the proper parameters for applications.

\subsection{Effects of different shutter and target thicknesses on ion acceleration in 2D} \label{sec:thickness_tar}
\begin{figure}[ht]
	\begin{center}
		\includegraphics[width=\linewidth]{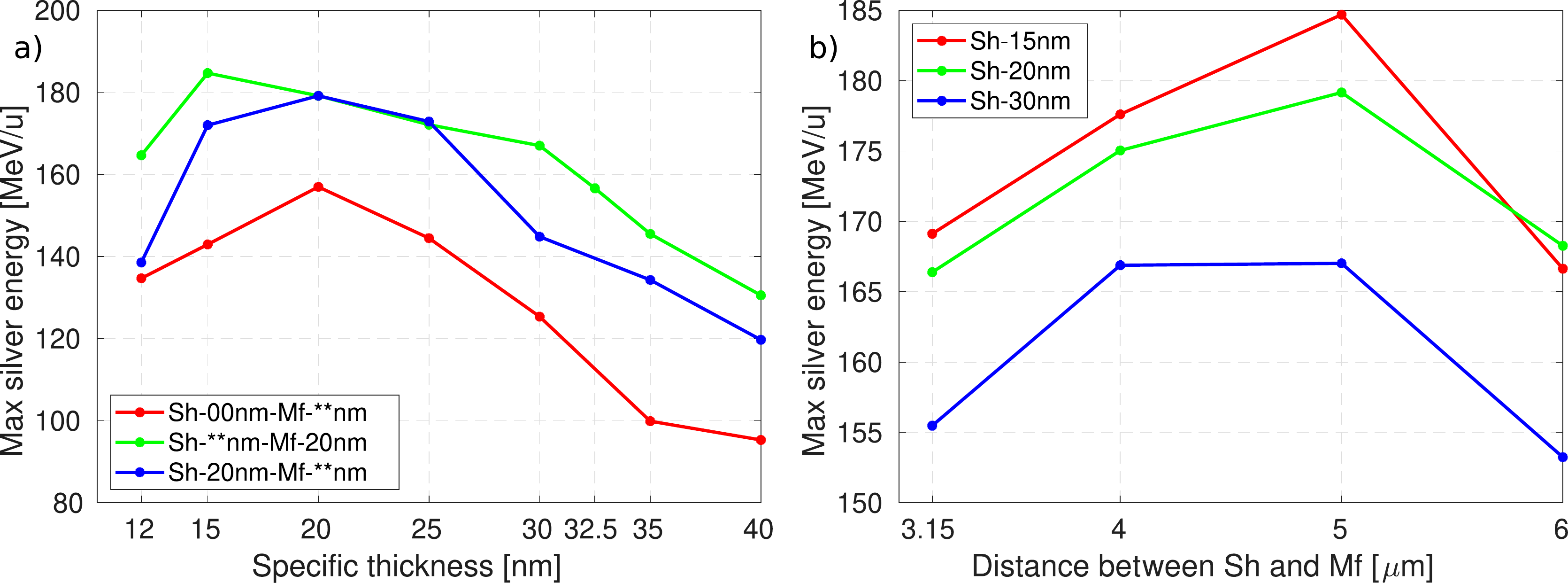}	
		\caption{\label{fig:ParamMfSh} (a) Influence of different combinations of shutter (Sh) and target - main foil (Mf) thicknesses on maximal ion energy. Red line - different thickness of the target without the plasma shutter; green line - different shutter thickness for 20 nm target; blue line - different target thickness for 20 nm shutter (b) Effect of different distance between the shutter and the target for different shutter thickness on maximal ion energy.}
	\end{center}
\end{figure}

For an efficient application of the transmitted laser pulse for ion acceleration, the benefits of the pulse shaping need to outweigh the loses of the laser pulse energy. Therefore, the combination of parameters of the plasma shutter (Sh) and the target - main foil (MF) like their thickness and distance between them are investigated in figure \ref{fig:ParamMfSh}. 

The optimal thickness for reaching maximal silver ion energy of the target without the use of the shutter is 20 nm as is shown by the red line in figure \ref{fig:ParamMfSh}(a). This value is slightly higher than the optimum predicted for RPA mechanism by Eq. \ref{eq:opt}, which corresponds to the thickness of 13 nm for the chosen silver target and laser pulse. The increase of the optimal target thickness can be ascribed to the early onset of the relativistic induced transparency in this ultrathin targets. Ion acceleration is then influenced by the hybrid RPA-TNSA mechanism described in \cite{Higginson2018_100MeV}. The next step is to compare various shutter thicknesses for the 20 nm target, which is shown by the green line. The maximal ion energy is increased by including the plasma shutter into the simulation, when its thickness is lower than 32.5 nm (for this cases the transmitted laser pulse energy is still above 50\% as discussed above). An interesting region for ion acceleration is found for shutter thicknesses between 15 nm and 30 nm where the maximal ion energy is slowly linearly decreasing with the shutter thickness. The maximum of 185 MeV/$A$ is reached in the case of 15 nm Sh. The thickness of 20 nm, where the maximal ion energy is just slightly lower (179 MeV/$A$) is chosen for further investigation as it can be more reliably represented in the 3D simulations with larger cells (keeping at least 4 cells per target in our simulations). The blue line in figure \ref{fig:ParamMfSh}(a) shows the dependence of maximal silver ion energy on the target thickness, when the shutter thickness is 20 nm. The maximum ion energy is reached for the same thickness of 20 nm as in the case without the Sh. Therefore, the plasma dynamics, including the onset of the relativistic induced transparency, is similar in both cases, as will be further discussed in section \ref{sec3D}. In addition, the ion acceleration is less sensitive to the target thickness between 15 nm and 25 nm when the shutter is included than in the corresponding case without the shutter. This can be ascribed to the steep-rising front generated by the plasma shutter reducing the pre-expansion of the target. So far the distance between the shutter and the target was set to 5 $ \mathrm{\mu m} $, which roughly corresponds to the length of the rising front in figure \ref{fig:Sh3D_slices}(b).  This distance is the optimum in our 2D cases as can be seen in figure \ref{fig:ParamMfSh}(b) comparing different distances. Moreover, a plateau of similar maximal energies between 4 and 5 $ \mathrm{\mu m} $ is developed for thicker plasma shutters, making the target fabrication more flexible. 

Regarding to the parametric scan above, the thicknesses used for 2D and 3D simulations of ion acceleration are hereinafter 20 nm for both shutter and target which are located 5~$ \mathrm{\mu m} $ apart. This set of parameters resulted in the increase of maximal ion energy from 157 MeV/$A$  to 179 MeV/$A$, when the plasma shutter is included. Note that similar but less significant increase was achieved with a lower-$ Z $ material for the same laser pulse and shutter parameters. Namely, using a fully ionized aluminum foil with thickness of 80 nm, the maximal aluminum ion energy increased from 230 MeV/$A$ to 243 MeV/$A$. Further lowering of $ Z $ of the target inherently results in increase of its optimal thickness according to eq. \ref{eq:opt}, which may alter the laser-target dynamics of ultrathin targets described in this work.

\subsection{Influence of the steep-rising front} \label{sec:Steep:front}
\begin{figure}[ht]
	\begin{center}
		\includegraphics[width=\linewidth]{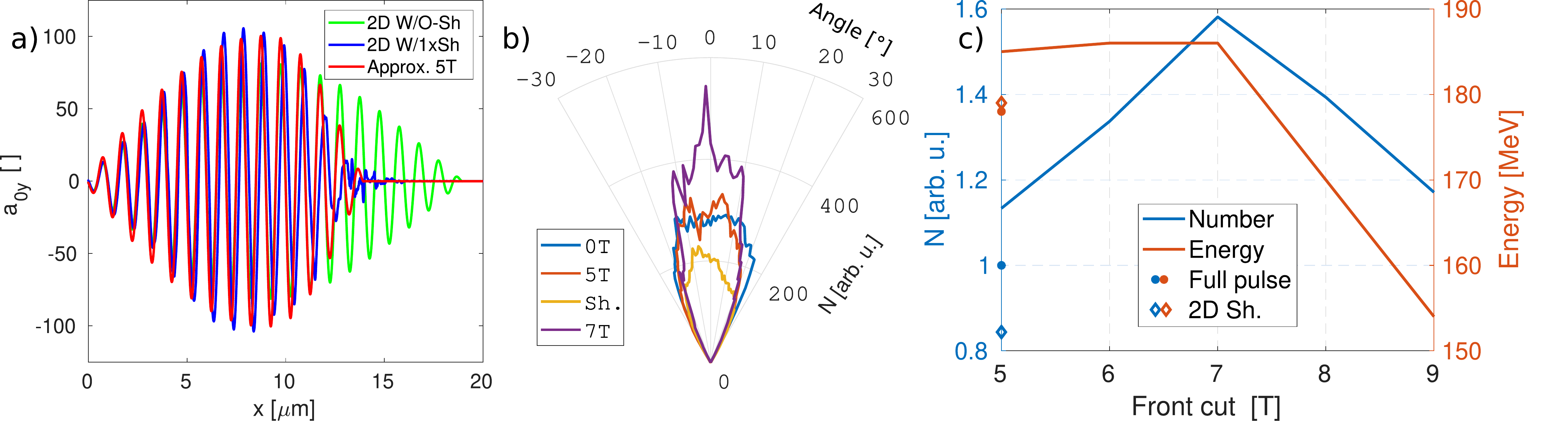}	
		\caption{\label{fig:SF} Influence of the steep-rising front of the laser pulse. (a) Comparison of the laser pulse transmitted through the plasma shutter (W/1xSh) with its approximation described in the text and the original pulse propagating in the vacuum (W/O-Sh) in 2D. (b) Angular distribution of the ions with energy over 90 MeV for different cutting of the pulse front, see the text for the description. (c) Dependence of maximal ion energy and number of ions (energy over 90 MeV) on the cutting of the pulse front.}
	\end{center}
\end{figure}

As the generation of the steep-rising front of the incident laser pulse is an important feature in the subsequent laser-target dynamics, we run several 2D simulations approximating the generated pulse shape. The approximated pulse assumed an increased intensity of the central 1D profile from $ a_0 =$ 85  to $ a_0 =$ 105, which fits well the 1D profile obtained from the 2D simulation as can be seen in figure \ref{fig:SF}(a).  The pulse then consists of two sinusoidal time envelopes. The beginning of the time profile is cut to zero for several periods (number of cut periods is denoted as Front cut \#T in figure \ref{fig:SF}) and then rises to maximal $ a_0 $ with sinusoidal shape till the half of the original pulse. This way the steep front is generated. The envelope of the second half of the laser pulse remains unperturbed and decreases from the maximum to zero at the same timescale as the original pulse.
 The spatial profile is kept the same as of the original pulse, ignoring the appearance of the local off-axis maxima reaching higher value of $ a_0= $ 127 (see the 2D slice from the 3D simulation in figure \ref{fig:Sh3D_slices}(a) for illustration). Thus, an average central on-axis maximum  is introduced for the sake of brevity. The comparison of the 1D profile generated by the plasma shutter (blue) with this approximation assuming the cut of 5~$ T $ (red) and the original pulse (green), at time when the pulse is transmitted through the plasma shutter and the 1D profile stabilized, is shown in figure \ref{fig:SF}(a). 

The introduction of the steep front has a positive effect on the divergence of high-energy ions (energy over 90 MeV/$A$, which is roughly half of the maximal energy reached). Their angular distribution for different front cutting is shown in figure \ref{fig:SF}(b). The use of the full pulse (0T - blue) results in a relatively broad flat-top
angular distribution without any significant peak. With the use of our steep-front approximation (cutting of 5T - red) the distribution is divided into two lateral peaks. Increasing the steepness of the laser pulse front even further (cutting of 7T - deep purple) results in a distinct single peak near the axis. The use of the plasma shutter in 2D (Sh. - yellow) resulted in lower number of particles, which can be ascribed to a different spatial profile, which is more localized compared to the Gaussian one, used in the approximations. Also the single peak generation is not so distinctive in this case in the 2D configuration. The benefits of the steep-rising front in the shutter case are more pronounced in 3D (see figure \ref{fig:Ion3D}(f) in the next section), where the intensity is higher, so the front is steeper even for a similar time scale and is then closer to the 7T case in 2D.
 
Another interesting effects of the steep-rising front is the increase of the number of high-energy ions in the region around the central axis ($y$ between $\pm3\ \mathrm{\mu m}$). This trend reaches maximum for the cutting of 7T, as can be seen in figure \ref{fig:SF}(c) by the blue line and blue $ y $-axis. Almost 1.6 more ions are located in this regions in the 7T case compared to the use of the full uncut pulse (denoted by the single point). After this distinct maximum the number of particles drops. As mentioned before, the number of particles in the shutter case (denoted by diamond) is lower due to different pulse representation in the space. The red line and $ y $-axis in figure \ref{fig:SF}(c) correspond to the maximal silver ion energy reached in the simulation. The development of the maximal energy is saturated till 7T and has the same drop as the number of particles  afterwards. These drops in the maximal energy and number of particles can explain the steeper reduce of the maximal energy of silver ions for green line in figure \ref{fig:ParamMfSh}(a) with shutter thickness increasing over 30 nm. The thicker plasma shutter cuts too significant part of the laser pulse, reducing the maximal ion energy. The energy reached by the case with the plasma shutter (red diamond) is similar to the one reached with the full pulse (red star). Therefore, the ion acceleration dynamics corresponds to the lower value of $ a_0 =105 $ from the on-axis 1D profile (figure \ref{fig:SF}(a)) and not to the off-axis local maxima of $ a_0 =127 $ (figure \ref{fig:ParamSh}(a))  observed in the previous chapters.    

\subsection{Ion acceleration in 3D} \label{sec3D}
\begin{figure}[ht]
	\begin{center}
		\includegraphics[width=\linewidth]{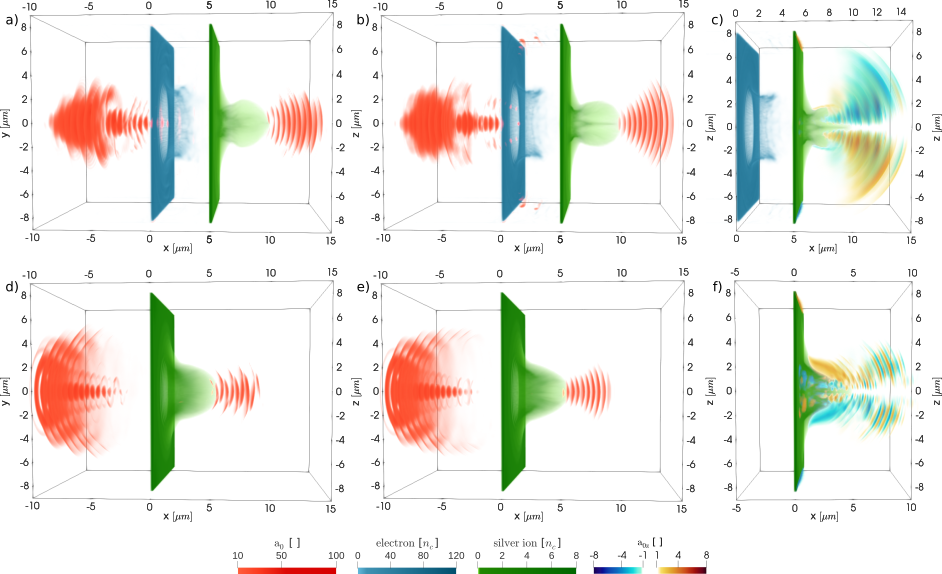}	
		\caption{\label{fig:BxSMari} Volumetric visualizations of the 3D simulations. First row contains the data from the simulation with the shutter and the second row without the shutter. (a,b,d,e) The laser pulse electric field ($ a_0 $)  is represented by the red scale, electron density by the blue scale and silver ion density by the green scale. (c,f) The generated electric field in the $ z $-direction represented by the blue to red scale is shown instead of the laser pulse.  The first column is the projection of $ x $-$ y $ plane, while the viewpoint in the second and third columns is rotated by $ 90^{\circ}$  into the projection of $ x $-$ z $ plane.  }
	\end{center}
\end{figure}

In this section, we will compare the silver target without the plasma shutter, placed at the laser focus at the $x$-position of 0 $ \mathrm{\mu m} $ and the silver target with the plasma shutter. In the later case the shutter is placed at the laser focus at $ x=0\  \mathrm{\mu m} $ and the silver target is located 5 $ \mathrm{\mu m} $ behind it. The simulation setup is shown as 3D volumetric visualization of laser pulse ($ a_0 $, red), plasma shutter (electron density, blue) and target (silver ion density, green) in figure \ref{fig:BxSMari}. The time instants are  $t=40$ $T$ in the case with the plasma shutter (first row) and corresponding $t=35 $ $T$ for the case without the shutter (second row) as the target is placed 5 $ \lambda $ closer to the beginning of the simulation box. The first column is the projection of $ x $-$ y $ plane, while the viewpoint in the second and third columns is rotated by $ 90^{\circ}$  into the projection of $ x $-$ z $ plane.
 
 As can be seen in the figure \ref{fig:BxSMari}(b), the application of the plasma shutter results into a narrow beam-like structure of accelerated silver ions (green) in the $ x $-$ z $ plane. This structure is hard to recognize in the case without the plasma shutter (figure \ref{fig:BxSMari}(e)) and is missing in the $ x $-$ y $ plane (figures \ref{fig:BxSMari}(a) and \ref{fig:BxSMari}(d)). The difference in the electric field in the $ z $-direction ($ a_{0z} $, blue-red scale) at the corresponding times is shown in the third column. In the case without the plasma shutter (figure \ref{fig:BxSMari}(f)) a relatively strong defocusing electric field is located around the outer surface of the ion cloud on positions $ y=\pm 2\ \mathrm{\mu m}$. The focusing field with opposite polarity is also present inside the ion cloud around position $x=3\  \mathrm{\mu m}$, $ y=\pm 0.5\ \mathrm{\mu m}$. However, its influence is limited only to the small region. When the plasma shutter is included (figure \ref{fig:BxSMari}(c)) the defocusing field acting on two lateral parts is reduced. On the contrary, the focusing field surrounding the central ion beam is well developed and prolonged over several microns.
 
\begin{figure}[ht]
	\begin{center}
		\includegraphics[width=\linewidth]{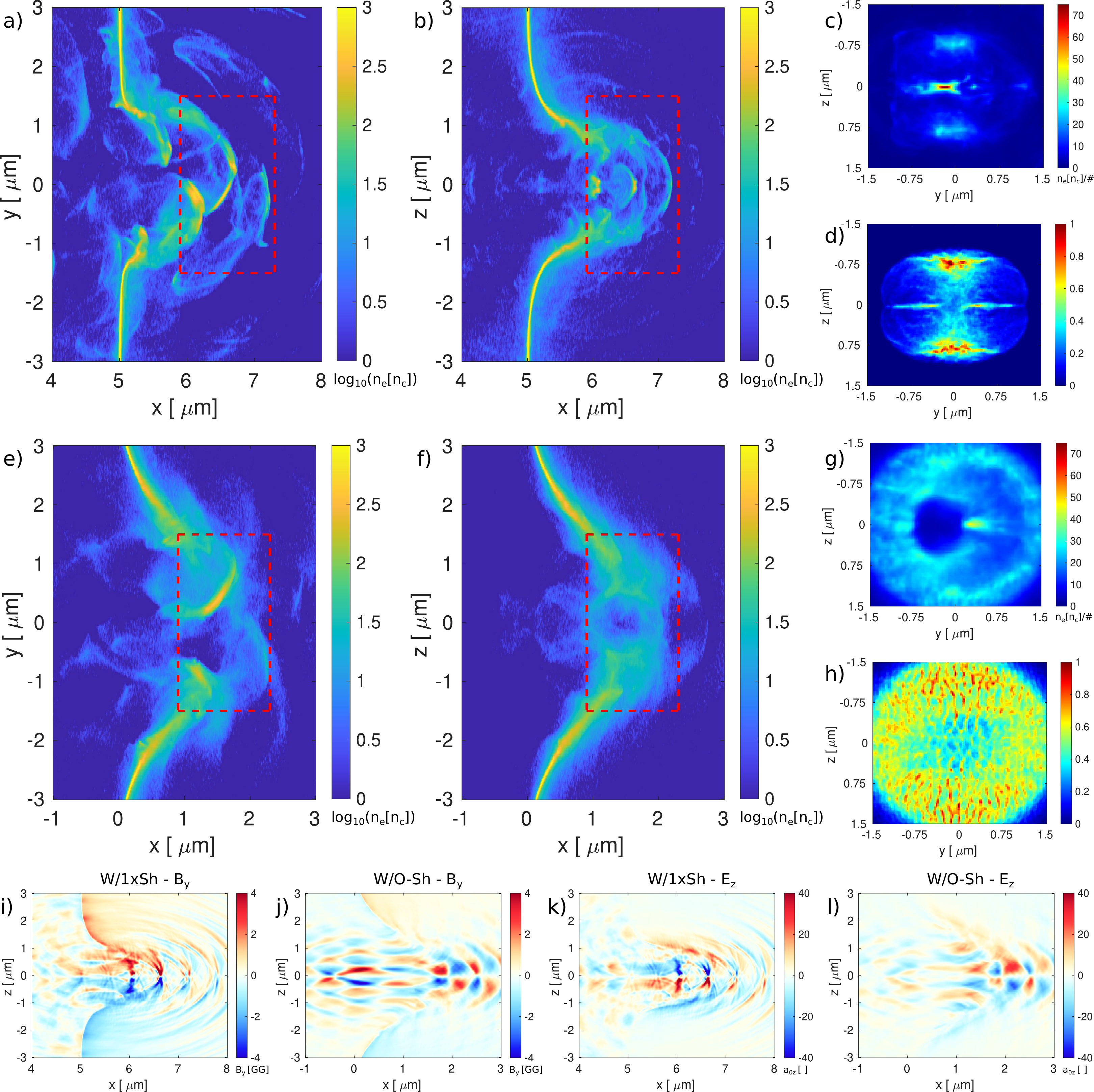}	
		\caption{\label{fig:Bunche} 2D slices from the 3D simulations. First row contains the data from the simulation with the shutter and the second row without the shutter. (a,b,e,f) Logarithm of the electron density in $ x $-$ y $ and $ x $-$ z $ planes. Average electron (c,g) and silver ion (d,h) density in $ y $-$ z $ plane (average of the grid points in the $ x $-direction in the volume denoted by the red rectangles in (a,b,e,f). The third row contains generated magnetic field in the $ y $-direction (i,j) and electric field in the $ z $-direction (k,l) from the simulations with the plasma shutter (W/1xSh) and without it (W/O-sh).}
	\end{center}
\end{figure}

The development of this heavy-ion beam-like structure is driven by the intense dynamics between the laser pulse and electron bunches at earlier stages, as can be seen in figure \ref{fig:Bunche}. The first row contains the visualization of the data from the case with the plasma shutter at time $t = 33$ $T$  and the second row contains the data from the case without the plasma shutter at the corresponding time $t = 28$ T.
 The third row contains generated magnetic and electric fields in the directions perpendicular to these of the laser pulse for both simulated cases.

In the $ x $-$ y $ plane (i.e., the polarization plane of the laser pulse) in figure \ref{fig:Bunche}(a,e), the high-density bunches appear at the alternating $ \pm y $ positions, roughly half of the laser pulse wavelength apart. The structure is more distinct in the case with the plasma shutter, where the electron density exceeds 400 $ n_c $. Moreover, the diameter of the final aperture is smaller compared to the case without the plasma shutter. The target outside the central region of $ y=\pm 1.5\ \mathrm{\mu m}$ remains unperturbed. The differences can be ascribed to the generation of the steep-rising front and increased intensity when the plasma shutter is included (as was shown in figure \ref{fig:Sh3D_slices}). Therefore, the mostly unperturbed foil can still interact with the high-intensity part of the laser pulse and the dynamics is more intense. On the contrary, in the case without the plasma shutter, the foil (and consequently the generated bunches) needs to interact with a relatively long low-intensity time profile of the incoming laser pulse, causing their pre-expansion, before the arrival of the high-intensity part. The same bunch structure is reflected in the $ x $-$ z $ plane in figure \ref{fig:Bunche}(b) around the position $ z=0 $. On the contrary, the structure is not observed in figure \ref{fig:Bunche}(f), as it is strongly pre-expanded. The figures \ref{fig:Bunche}(c,g)  show the average electron density (its summation divided by the number of grid points in the $ x $-direction) in the $ y $-$ z $ plane over the same volume denoted by the red rectangles in figures \ref{fig:Bunche}(a,b,e,f). Three distinct regions are presented in figure \ref{fig:Bunche}(c), two lobs around $ z=\pm0.75\ \mathrm{\mu m} $ corresponding to the aperture diameter and a distinctive high-density lob around $ z=0\ \mathrm{\mu m} $. The lobs are prolonged in the $ y $-direction, which corresponds to the spread by the laser pulse polarization. The lateral lobs were observed in Ref. \cite{Gonzales2016electrons} using a single foil of a lighter material, whereas the amount of electrons in the central lobe was strongly limited in that case. A less distinctive central lob with lower density is also present in the case without the plasma shutter (figure \ref{fig:Bunche}(g)). The effect of laser polarization on lateral lobs is weaker in this case, as they complete a ring shape of a similar density around the central lob.

The electron bunches in the case with the plasma shutter are also well coupled with the corresponding magnetic field in the $ y $-direction ($ B_y $) as can be seen in figure \ref{fig:Bunche}(i). The field exceeds 4 gigagauss at this timeframe (colors are saturated). This clear structure is not present in the case without the plasma shutter in figure \ref{fig:Bunche}(d), corresponding to the different electron distribution. The structuring of generated magnetic and electric fields using thin targets becomes a subject of interest recently, e.g., by using flat channel-like targets for improvement of laser-driven particle parameters like beam divergence \cite{Zakova_2021_quadrupole}.  

The differences in the electron distribution then manifests itself also in the corresponding ion distributions in figures \ref{fig:Bunche}(d) and \ref{fig:Bunche}(h) via the generated electric field in the $ z $-direction, which are shown in figure \ref{fig:Bunche}(k) and \ref{fig:Bunche}(l). In the case with the plasma shutter (figure \ref{fig:Bunche}(k)), two regions of electric fields with opposite influence on ions arises: 1) the parabolic shape structure which corresponds to the outer electron cloud around $ y=\pm 0.75\ \mathrm{\mu m}$ in figure \ref{fig:Bunche}(b) and 2) the region inside this cloud with opposite polarity of the electric field.  In this inner region, the Coulomb force $ F=qE $ focuses ions (with positive charge $ q $) towards $ z=0 $ as the electric field is negative for $ z>0 $ and positive for $ z<0 $. This field is present from the $ x\approx5.5\ \mathrm{\mu m} $ (closely behind the initial target position at $ x=5\ \mathrm{\mu m} $) to the peak of the parabolic structure around $ x\approx7.5\ \mathrm{\mu m} $. The field (figure \ref{fig:Bunche}(k)) is stronger at the positions corresponding to the electron bunches in figure \ref{fig:Bunche}(b). Note that the colors are saturated for visualization purposes, the peak field reaches locally values of $ a_{0z} \approx \pm 120 $, even higher than the initial laser pulse with $ a_{0y} = 85 $. On the contrary, the parabolic structure with opposite polarity produces a de-focusing field pushing ions away from the central axis. The structures in electric field then correspond to the ion profile in the $ y$-$z $ plane in figure \ref{fig:Bunche}(d) with two ion lobs pushed away from the central axis to positions $ y=\pm 0.75\ \mathrm{\mu m}$ and two thin ion stripes around $ z=0 $, which follows the electron bunches in the $ y $-direction (figure \ref{fig:Bunche}(a)).
   
The electric field in the case without the plasma shutter (figure \ref{fig:Bunche}(l)) contains similar de-focusing parabolic structure corresponding to a broader electron cloud in \ref{fig:Bunche}(f). On the contrary, the inner structure behind the target till $ x\approx1.5\ \mathrm{\mu m} $ has opposite polarity around $ z=0 $ than in the case with the plasma shutter and is thus de-focusing. It results in the low-density region around center in the corresponding ion distribution \ref{fig:Bunche}(h)). The ion distribution also contains the transverse instability. The typical longitudinal stripes along $ z $-axis observed in the similar figures in 3D simulations and experiments for linear polarization in Refs. \cite{Gonzalez2016proton,Sgattoni2015RTI,Palmer2012RTI} are not so clearly visible in figure \ref{fig:Bunche}(h), as the inner field is weaker than the one of the parabolic structure (figure \ref{fig:Bunche}(l)). On the contrary, the inner field is the dominant part in the case with the plasma shutter (figure \ref{fig:Bunche}(k)), producing the stripe in the middle of figure \ref{fig:Bunche}(d). Therefore, the transverse instability is suppressed, when the plasma shutter is included, as can be seen by comparing the figures \ref{fig:Bunche}(d) and \ref{fig:Bunche}(h). This observation is in agreement with Ref. \cite{Matys2020}, where the steep-front laser pulse (which was assumed to be possibly produced by the plasma shutter) was used for mitigation of this kind of instabilities.

\begin{figure}[ht]
	\begin{center}
		\includegraphics[width=\linewidth]{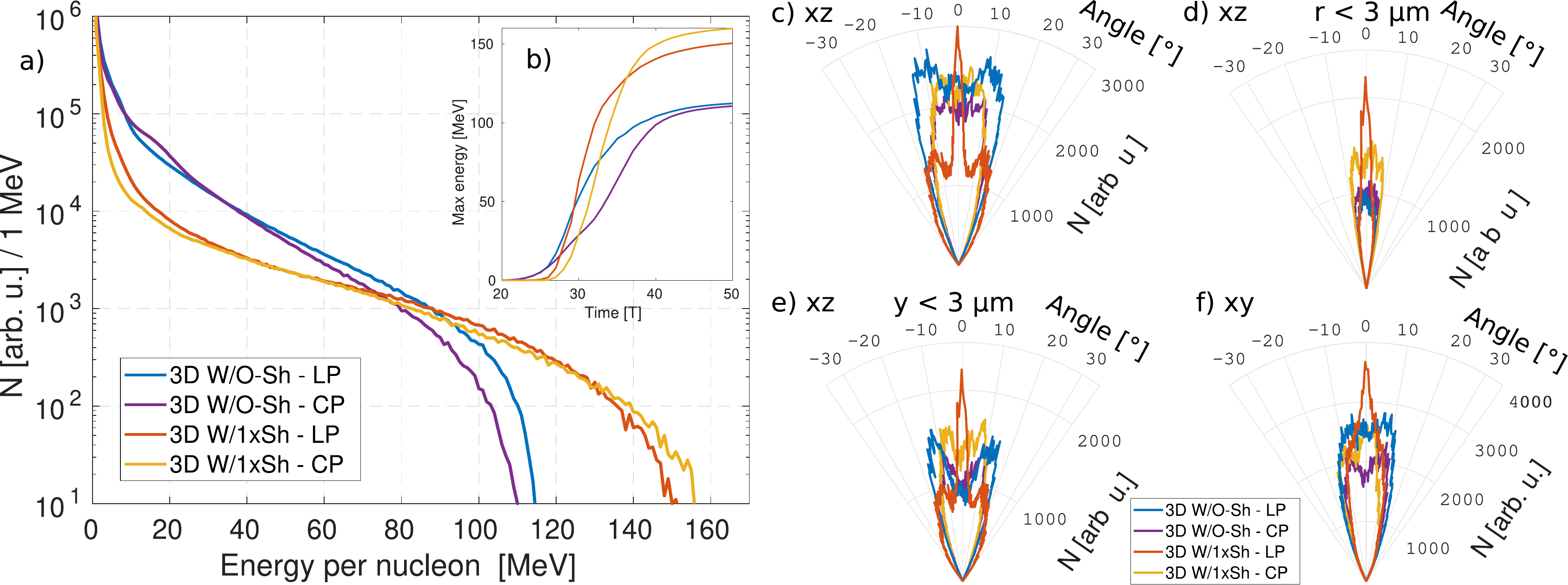}	
		\caption{\label{fig:Ion3D} Properties of silver ions from the 3D simulations with the plasma shutter (W/1xSh) and without (W/O-sh)  for linear (LP) and circular (CP) polarization. (a) Energy spectra at the end of the simulation ($ t=70\ T $), (b) time evolution of the maximal energy. Angular distributions of ions with energy over 55 MeV/$A$ in the $ x $-$ z $ plane (c,d,e) and $ x $-$ y $ plane (f).}
	\end{center}
\end{figure}

This development affects the final silver ion energy spectra and angular distribution as is shown in figure \ref{fig:Ion3D} at time $ t = 70\ T  $, when the acceleration in all cases already ended. 

The circular polarization (CP) is often proposed for the RPA dominated regimes of ion acceleration as an alternative to linear polarization (LP) to increase the maximal ion energy, reduce energy spread and divergence of the ion beam \cite{Henig_Tajima_2009,Chen2009,Klimo2008PRST, Robinson2008}. Therefore, additional 3D simulations with and without the plasma shutter using CP were made for comparison. Their results are also included in figure \ref{fig:Ion3D} via yellow and deep purple lines and discussed in the last paragraph of this section.

 In the LP cases, maximal energy rises from 115 MeV/$A$ to 155 MeV/$A$ (about 35\%) when the plasma shutter is included in the simulation. The increase occurs, even though about 50 \% of the laser pulse energy is consumed by the plasma shutter as was shown in figure \ref{fig:ParamSh}(b). The integrated number of high-energy ions is larger in the simulation with the shutter for energy above 65 MeV/$A $, although the total number of all accelerated ions is lower. The time evolution of the maximal energy of silver ions are shown in the inset of figure \ref{fig:Ion3D}(b). The time was shifted by 5 $ T $ in the case without the plasma shutter as the target is shifted by 5 $\lambda $. The time profile is similar in both cases. The gradual change from the  exponential rise, corresponding to the radiation pressure acceleration \cite{Esirkepov2004} (which can be in principle  unlimited \cite{Bulanov2010Unlim}), to logarithmic rise can be spotted in both lines around the time $ t = 30\ T$. The target becomes partially (relativistically) transparent around this time, as can be inferred from the figure \ref{fig:BxSMari}(e). The transmitted laser pulse front is about 10  $ \mathrm{\mu m} $ ahead from the target at time $ t= 40\ T$.  The RPA still significantly contributes to ion acceleration till time 33 $ T $ in the simulation with the shutter as the front of electron layer around the position $ x = 7\ \mathrm{\mu m} $ (figure \ref{fig:Bunche}(b)) is still compact. However, the density of the electron layer is decreasing, resulting in slower rise of maximal energy and partial transparency. Therefore, other mechanisms involving the relativistic transparency, like hybrid RPA-TNSA \cite{Higginson2018_100MeV} and directed coulomb explosion \cite{Bulanov_SS_DCE} also take place at this stage. The rise of maximal energy in this stage (33 $ T $ -- 40 $ T $) is noticeably lower compared to the previous stages. At time 40 \textit{T} most of the laser pulse already left the area where the plasma is located as can be seen in figure \ref{fig:BxSMari}(b). Only a small rise of ion energy is observed afterwards. Details on the influence of different acceleration mechanisms, supported by density profiles of electrons and silver ions, are included in the appendix.

The beam-like structure is reflected in the angular distribution of accelerated ions (energy above 55 MeV/$A$) in figures \ref{fig:Ion3D}(c--f) at time $ t = 70\ T  $.  In the $ x $-$ z $ plane (comparing momenta in $ x $ and $ z $ directions) in figure \ref{fig:Ion3D}(c), a narrow central bunch of particles is visible  when the shutter is included (red). The divergence at the FWHM of the bunch is around $ 5^{\circ}$, whereas the case without the shutter generates a broad bunch divided into three directions with overall divergence over $ 35^{\circ}$. Another advantage of the use of the shutter is the space positioning of the ion bunch. In the case without the plasma shutter, the central part of the ion angular distribution is significantly reduced when a "pinhole" is introduced filtering out the ions with radial position $r >  3 $ $ \mathrm{\mu m} $ away from the central axis, as can be seen in figure \ref{fig:Ion3D}(d). If the filtering is done only in the y-direction, the lateral parts of the bunch prevails over the central part in the case without shutter (figure \ref{fig:Ion3D}(e)). On the contrary, the central part prevails in the case with the shutter. The shutter has also a positive effect on the divergence at the FWHM in the $ x $-$ y $ plane \ref{fig:Ion3D}(f)), which is reduced from over $ 28.5^{\circ}$  to  about $ 18.5^{\circ}$. This can be ascribed to the generation of the steep-rising front of the incident laser pulse as is discussed in the section \ref{sec:Steep:front}. 

\begin{figure}[ht]
	\begin{center}
		\includegraphics[width=\linewidth]{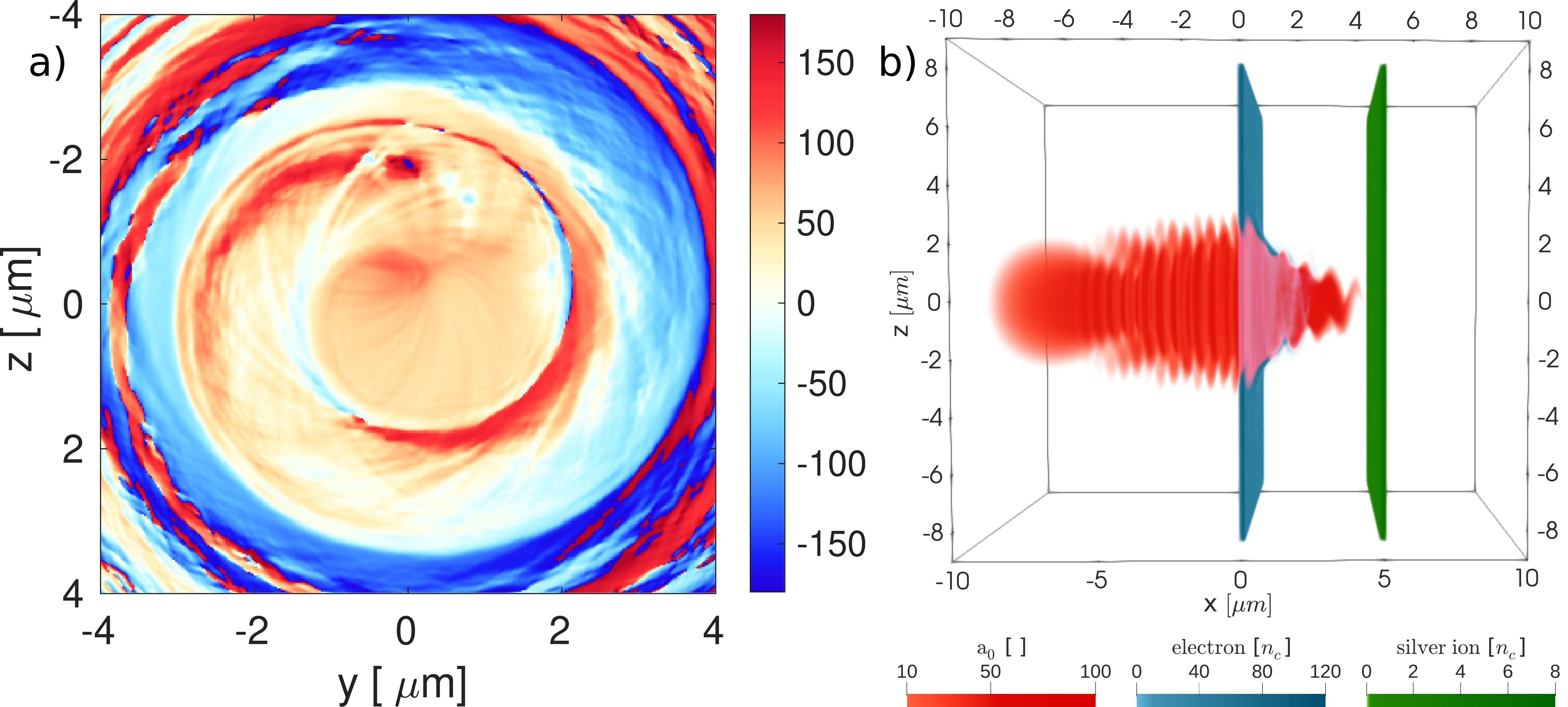}	
		\caption{\label{fig:Spiral} Spiral structure  in the 3D simulations with circular polarization. (a) The angle between $ E_y $ and $ E_z $ fields (calculated as four-quadrant inverse tangent) in the plane $ y $-$ z $, at position $ x = 5\ \mathrm{\mu m} $ from the simulation with only shutter (the target was not included in this simulation. This angle is constant in individual $ y $-$ z $ planes for the circular polarization before penetrating the plasma. b) 3D volumetric visualization from the simulation with both the shutter and the target (3D W/1xSh - CP) at earlier time. The laser pulse electric field ($ a_0 $)  is represented by the red scale, electron density by the blue scale and silver ion density by the green scale.}
	\end{center}
\end{figure}

In the CP cases, the silver ion spectra are similar to their LP counterparts (figure \ref{fig:Ion3D}(a)).  The increase of maximal ion energy when the plasma shutter is included is even slightly higher than for the linear polarization (about 44 \%). In the simulations with the shutter, the change of the polarization affects mostly the tail of the silver ion spectra, which experiences slight increase of maximal energy from 155 MeV/$A$ to 164 MeV/$A$.  The time evolution of maximal ion energy of the CP simulations (figure \ref{fig:Ion3D}(b)) infers that the main acceleration phase starts later and lasts longer than in the LP cases. The CP inherently stabilizes the pulse interaction with the plasma shutter (and the main foil) and thus delays the onset of relativistic transparency.  It also results in lower amount of pulse energy being transmitted  through the plasma shutter (38.4\% compared to  47.9 \% for the LP). On the other hand, the effect of divergence reduction with the inclusion of the shutter is strongly reduced in the CP simulations. The divergence of the silver ion beam is higher compared to the LP simulation with the shutter (but still lower compared to the LP case without the shutter) as can be seen in figures \ref{fig:Ion3D}(c-f)). In the simulations with the shutter the divergence increases from $5^{\circ}$ to about $27^{\circ}$ in the $ x $-$ z $ plane and from $ 18.5^{\circ}$ to about $ 25^{\circ}$ in the $ x $-$ y $ plane compared to the LP case.  Note that the CP laser pulse transmitted through the shutter also generates a spiral-like, rotating laser diffraction structure as previously observed in Refs \cite{Gonzales2016electrons,Bulanov_2000_spiral}. The spiral structure from our 3D simulations is shown in figure \ref{fig:Spiral}.

\subsection{The use of a double plasma shutter and prepulse filtering}
\label{sec:DoublePSh}

The interaction of the ultraintense pulse with the target having a step-like density profile may be hard to achieve even with nowadays technology. Therefore, we propose the use of a second plasma shutter, which will be used to mitigate the prepulses accompanying the main pulse.

\begin{figure}[ht]
	\begin{center}
		\includegraphics[width=\linewidth]{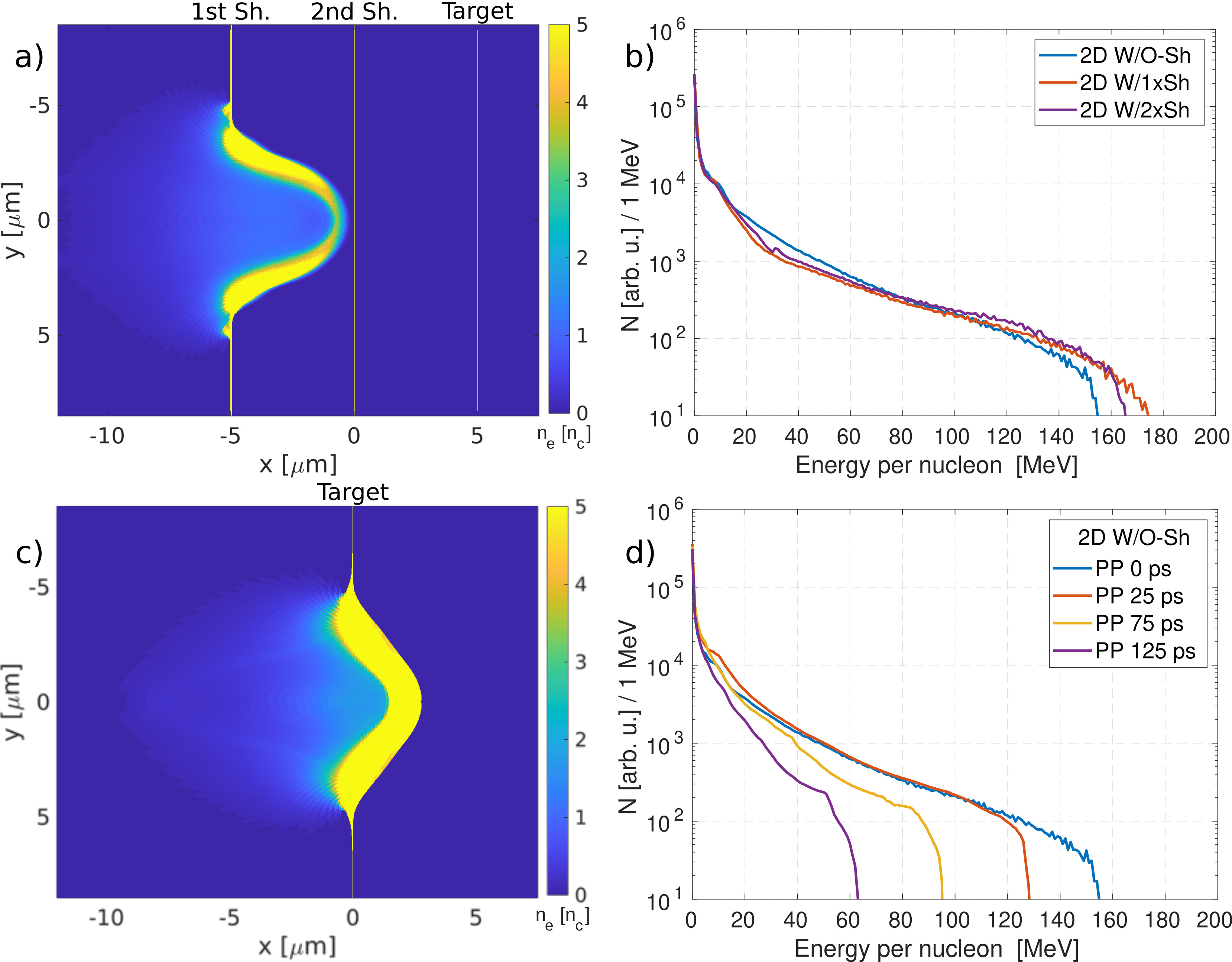}	
		\caption{\label{fig:DoublePS} Combining PIC and hydrodynamic simulations. Electron density at the beginning of the PIC simulations: (a) using the double plasma shutter (first one pre-expanded by the 125 ps long prepulse, silver target at $ x=5\ \mathrm{\mu m} $). (c) without any plasma shutter (silver target at $ x=0\ \mathrm{\mu m} $, pre-expanded by the same prepulse).   Colors are saturated at $ n_e=5\ n_c $. Silver ion energy spectra at the end of the simulations ($ t=70\ T $): (b) Comparing the 2D cases from the section \ref{sec:thickness_tar} without the shutter (W/O-sh), and with the shutter (W/1xSh) with the simulation using the double plasma shutter (W/2xSh). (d) The use of different length of prepulses (PP) in the simulations without the shutter.}
	\end{center}
\end{figure}

We run a 2D hydrodynamic simulation of prepulse with intensity of $10^{12}\, \mathrm{W}/\mathrm{cm}^2$ interacting with an aluminum shutter with areal density roughly the same as in the case of silicon nitride shutter. The 2D density profile from this simulation after 125 ps of interaction is then used as input data into a 2D particle-in-cell simulation in a form of a pre-expanded shutter, placed at the position $ x = -5\ \mathrm{\mu m}$. The simulation also contains the previously used setup of the non-expanded plasma shutter at $ x = 0\ \mathrm{\mu m}$ and silver target at $ x = 5\  \mathrm{\mu m}$ (see figure \ref{fig:DoublePS}(a)). This case is hereinafter referred to as W/2$\times$Sh in analogy to the reference 20 nm cases without any shutter (W/O-Sh) and with one non-expanded shutter (W/1xSh) used in previous sections.  The remaining electron density of the expanded plasma shutter is still above the critical density. Therefore, the plasma shutter can efficiently filter out this kind of prepulses with duration (at least) up to 125 ps, justifying the step-like densities of the second plasma shutter and the silver target in the simulation of this case. 

The silver ion energy spectra at the end of the simulation ($ t= 70\ T $) of the cases W/O-Sh (blue), W/1$\times$Sh (red) and W/2$\times$Sh (deep purple) from 2D simulations are shown in figure \ref{fig:DoublePS}(b).  When both shutters are used in the simulation, the maximum silver energy still increases from 157 MeV/$A$ to 167 MeV/$A$ compared to the simulation without the shutter. Note that the energy spectra of the cases of W/1$\times$Sh and W/2$\times$Sh are very similar for ions with energy up to 160 MeV/$A$ and only the most energetic ones are affected by the addition of the pre-expanded shutter. Therefore, maximal energy closer to the case of W/1$\times$Sh (179 MeV/$A$) can be theoretically reached with optimal thickness, expansion and positioning of the pre-expanded shutter. The 2D simulations in figure \ref{fig:DoublePS}(b) results in higher maximal energies than their 3D counterparts in figure \ref{fig:Ion3D}(a). The simulations without shutter experience more significant drop of the maximal energies when the simulation dimension increases from 2D to 3D (42 MeV/$A$) compared to the simulations with the shutter (24 MeV/$A$). Therefore, the increase of the maximal energy in the W/2$\times$Sh case compared to  W/O-Sh case may be more significant in 3D even for the presented configuration.

 One also needs to keep in mind, that the ion energy in the case W/O-Sh are overestimated compared to the W/2$\times$Sh case, as no prepulse was assumed in this simulation. When the same prepulse of 125 ps is used in the hydrodynamic simulation with the silver target, the originally ultra-thin target of 20 nm expanded into a few microns of overdense plasma predeceased by a long preplasma (figure \ref{fig:DoublePS}(c)). It results in a significant drop of the maximal energy compared to the previously optimized simulation. Several simulations with the density profiles obtained from the different time of hydrodynamic simulation with the prepulse were performed. The decrease of the maximal energy with the increase of the prepulse duration (PP) is shown in figure \ref{fig:DoublePS}(d). Therefore, to fully understand the impact of the double plasma shutter scenario, one needs to compare the deep purple lines in figure \ref{fig:DoublePS}(b) and \ref{fig:DoublePS}(d), which captures the same physics. This comparison provides the increase of maximal energy from 64~MeV/$A$ to 167 MeV/$A$.  
  
\begin{figure}[ht]
 	\begin{center}
 		\includegraphics[width=0.75\linewidth]{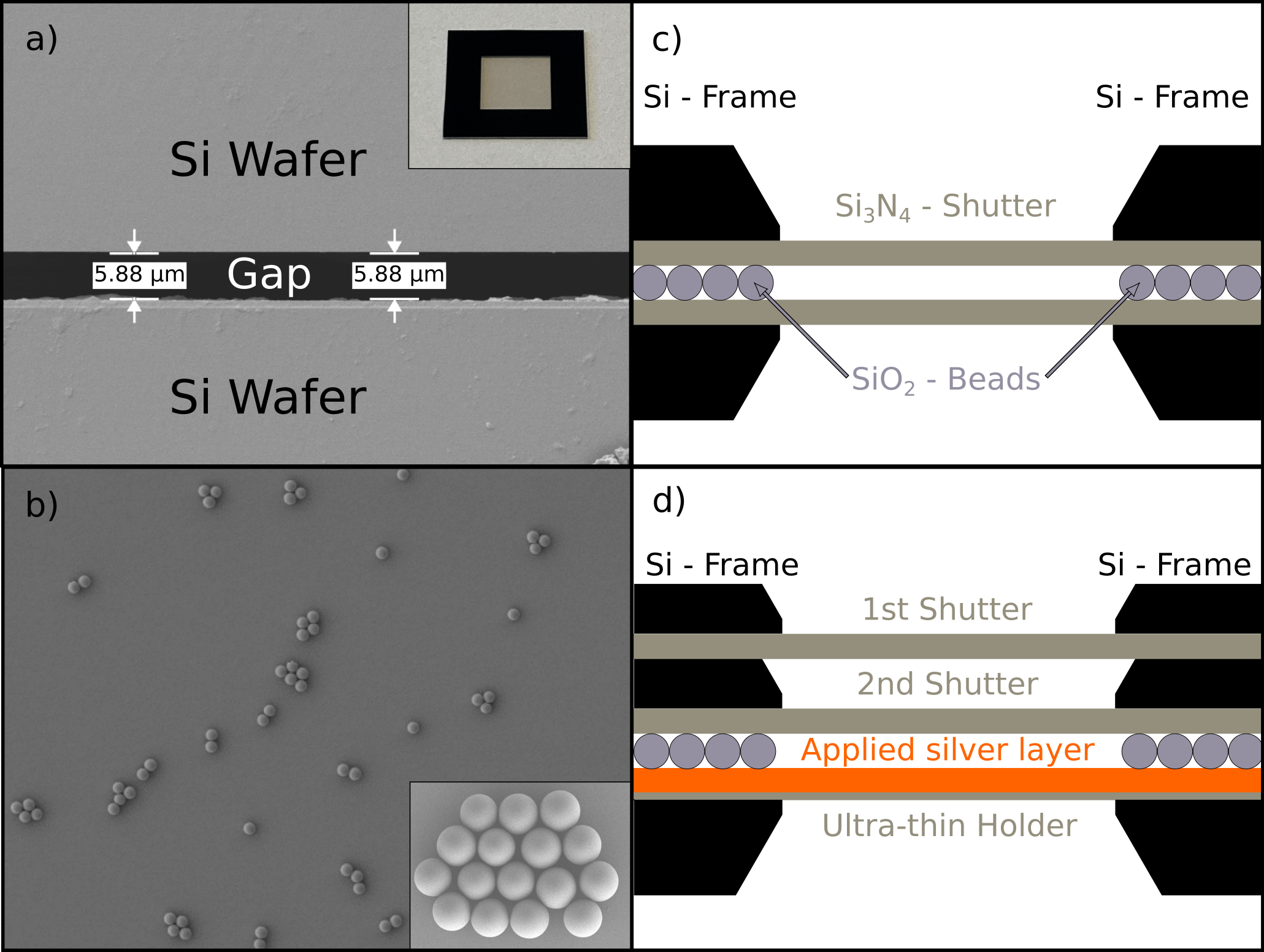}	
 		\caption{\label{fig:Prototype} Prototype of the double plasma shutter. (a) SEM micrography - side view on the gap between two plane-parallel silicon frames. The top view of membrane with silicon frame is shown in the inset, window size is 5x5 mm. (b) SEM micrography - top overview of a sililicon surface with the micro-islands created by closely packed SiO$_2 $ microspheres after drying up of a droplet of ethanolic dispersion. A partially enlarged detail image is embedded in the lower right corner. The diameter of microspheres is 5.7 $ \mathrm{\mu m} $ with standard deviation of 0.2 $ \mathrm{\mu m} $ (c) Schematic cross-section depicting arrangement of the sandwich-like assembly of two windows with microspheres as bondline spacers to maintain a controlled gap.  (d) Schematic cross-section of the double shutter realized by two windows oriented in the same direction and target attached with the similar set-up using microspheres as in c).}
 	\end{center}
 \end{figure}

 Our theoretical finding leads to the formulation of an idea of a double plasma shutter, which prototype was prepared at the Czech Technical University (see the photo in the inset of figure \ref{fig:Prototype}(a)).
  It consists of two commercial silicon nitride membranes separated by $ \mathrm{SiO_2} $ monodisperse microspheres. Therefore, a parallel surface between two layers is provided with the same spacing of 5.88 $ \mathrm{\mu m} $ as can be seen in the snapshot from the scanning electron microscope (SEM) in figure \ref{fig:Prototype}(a). The two membranes have a thick protective frame made of silicon (thickness of $200\ \mathrm{\mu m} $) with a thin window of silicon nitride (30 nm). The window is located at the edge of the one side of the membrane. Therefore, the double plasma shutter is created by these two windows facing each other as depicted in the scheme in figure \ref{fig:Prototype}(c). Multiple windows can be present on one shutter. Therefore, it can be used for high-repetition experiments using the target tower \cite{Margarone2018_Elimaia,Chagovets2021_targetry_automation}. For our laser and target parameters (requiring only a small gap between the shutter and target) another approach, with the two shutter windows oriented in the same direction (as shown in figure \ref{fig:Prototype}(d)), may be more practical. Note that for this implementation the manufactured frame of the shutter window should be thinner than in figure \ref{fig:Prototype}(c). The target can be attached either as the third window oriented in the same direction with the frame thickness corresponding to the required gap, using the microspheres (as in the previous scheme) or other spacer elements attached directly by the manufacturer. The microspheres option (shown in figure \ref{fig:Prototype}(d)) provides a possibility of an on-site modification of the gap size and the target structure for several experimental setups with prefabricated double shutter. The silver target can be either a standalone silver foil or a silver layer applied to an ultra-thin silicon nitride holder. The holder can be fabricated with thickness down to a few nm \cite{Yanagi_3nm} to limit its effects on the silver ion acceleration. Note that the silver layer can be applied to either side of the holder, depending on the application.

\section{Conclusion} \label{sec:Conclusion}
In conclusion, we investigate the application of the plasma shutter for heavy ion acceleration driven by a high-intensity laser pulse using 3D and 2D PIC simulations. 
The laser pulse, transmitted through the plasma shutter, gains a steep-rising front and its peak intensity is locally increased at the cost of losing part of its energy, depending on the shutter thickness. These effects have a direct influence on subsequent ion acceleration from the ultrathin target located behind the plasma shutter. The parameters like target and shutter thicknesses, size of the gap between them, and the effects of the pulse-front steepness are investigated using 2D PIC simulations for the case of a silicon nitride plasma shutter, a silver target and a linearly polarized laser pulse. In the follow-up 3D simulations of our reference case, the maximal energy of silver ions increases by 35\% when the plasma shutter is included. Moreover, the steep-rising front of the laser pulse leads to the formation of high density electron bunches. This structure (which also appears in the generated electric and magnetic fields) focuses ions towards the laser axis in the plane perpendicular to the laser polarization. The generated high energy ion beam has significantly lower divergence compared to the broad ion cloud, generated without the shutter. In the absence of the plasma shutter, the structures are pre-expanded by the low intensity part of the laser pulse and the subsequent onset of transverse instability. This instability observed in ion density is thus efficiently reduced using plasma shutter. 3D simulations with the circular polarization with  and without the shutter are performed for comparison.  The increase of maximal ion energy in the simulation with the plasma shutter is observed also for the circular polarization (by 44~\%). The use of the circular polarization in combination with the plasma shutter results in a slight increase of maximal ion energy, but also increase of the ion beam divergence compared to the corresponding simulation with a linear polarization. 

The efficiency of the processes introduced by the plasma shutter may be reduced by the previous shutter interaction with a long prepulse. 
Therefore, the effects of sub-ns prepulses are investigated using a combination of 2D PIC and hydrodynamic simulations assuming a double plasma shutter. The first shutter can withstand the assumed sub-ns prepulse (treatment of ns and ps prepulses by other techniques is assumed, alternatively increasing the  thickness of the first shutter may filter out longer prepulses). Therefore, the processes of the steep front generation and the local intensity increase can develop via interaction with the second non-expanded shutter. The increase of the maximal ion energy compared to the simulation with a step-like density target without any shutter is demonstrated also in this case.  Moreover, the comparison with a silver target pre-expanded by the same sub-ns prepulse as in the double shutter scenario results in an increase of maximal silver energy by the factor of 2.6 in the 2D simulations. A prototype of this double shutter is presented and the design of the whole shutter-target setup is discussed. 

\ack
Portions  of  this  research  were  carried  out  at  ELI  Beamlines,  a 
European  user  facility  operated  by  the  Institute  of  Physics  of  the  Academy  of  Sciences  of  the  Czech Republic.
Our work is supported by projects: High Field Initiative (CZ.02.1.01/0.0/0.0/15\_003/0000449) and Center of Advanced Applied Sciences (CZ.02.1.01/0.0/0.0/16\_019/0000778) from the European Regional Development Fund.

This work was supported by the Ministry of Education, Youth and Sports of the Czech Republic through the e-INFRA CZ (ID:90140). Access to CESNET storage facilities provided by the project e-INFRA CZ under the programme Projects of Large Research, Development, and Innovations Infrastructures (LM2018140), is appreciated. The data post-processing was performed using computational resources funded from the CAAS project. 

The support of Grant Agency of the Czech Technical University in Prague is appreciated, grants no. SGS22/185/OHK4/3T/14 and SGS22/184/OHK4/3T/14.

 We appreciate the collaboration with Lucie Maresova from the Czech Technical University in Prague taking the snapshots of the double plasma shutter via SEM micrography and with Virtual Beamline team of ELI Beamlines Centre, namely P.~Janecka and J.~Grosz with the work done towards the virtual reality visualization of our results. Fruitful discussions with M. Greplova-Zakova from ELI Beamlines Centre, K. Mima from Institute of Laser Engineering, Osaka University, Japan and E. Daigle and H. Hosseinkhannazer from Norcada Inc., Edmonton, Canada are gratefully acknowledged.

\appendix
\section*{Appendix A: Details on the influence of different acceleration mechanisms}
\begin{figure}[ht]
	\begin{center}
		\includegraphics[width=\linewidth]{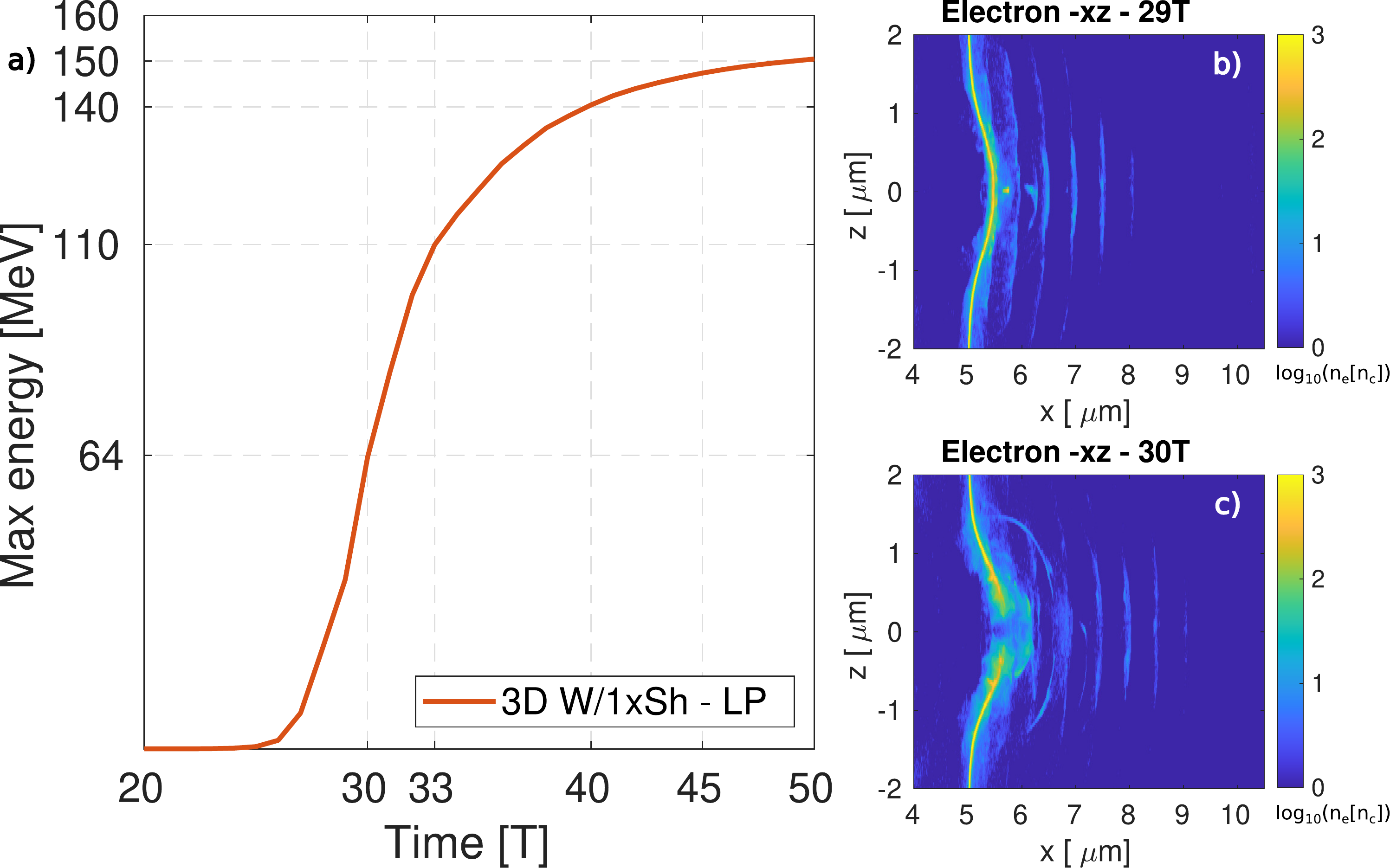}	
		\caption{\label{fig:App1}  a) Time evolution of maximal ion energy. b,c) Electron density around the time of laser penetration from the 3D simulation with the shutter and linear polarization (3D W/1xSh - LP).}
	\end{center}
\end{figure}

\begin{figure}[ht]
	\begin{center}
		\includegraphics[width=\linewidth]{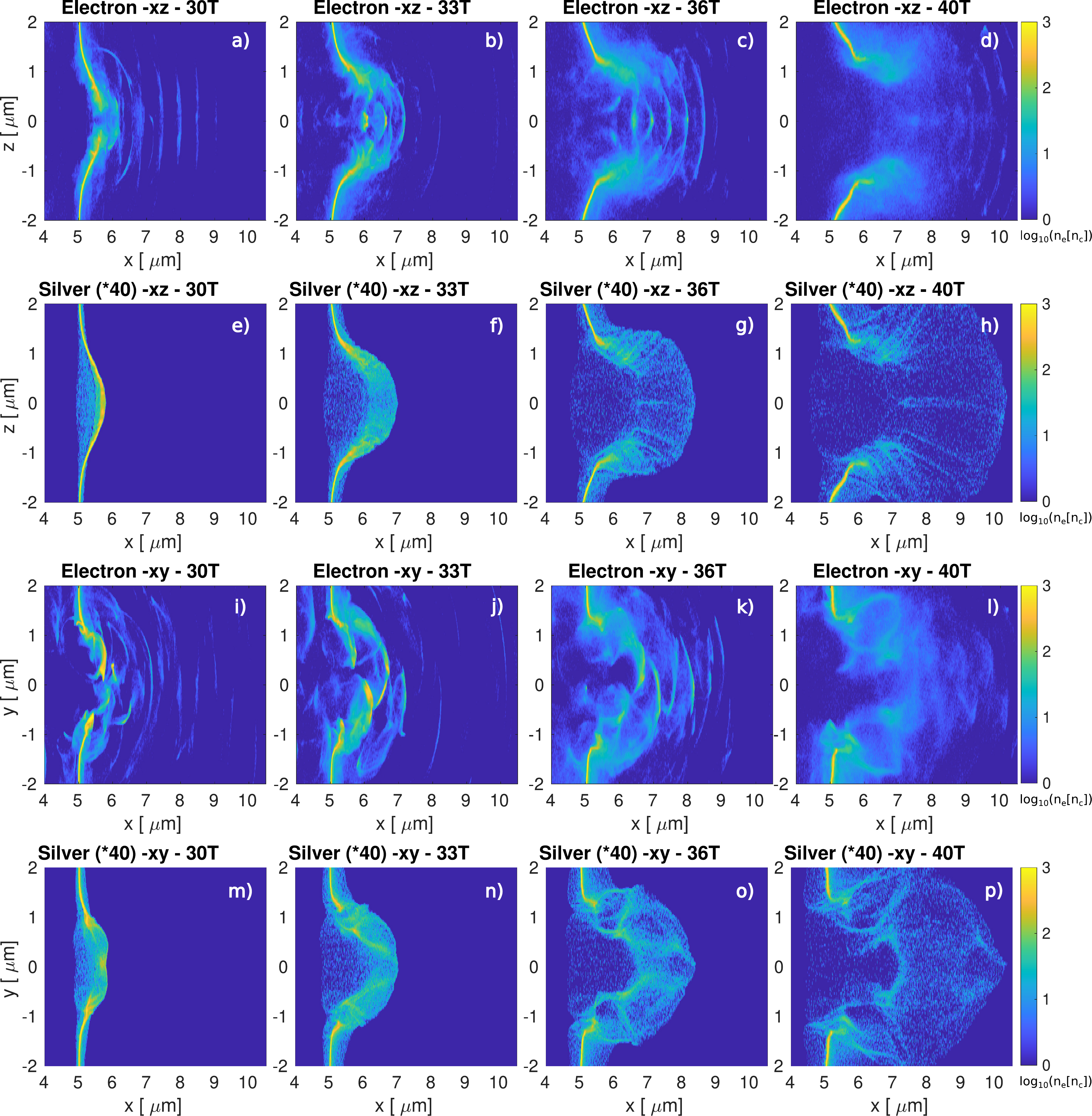}	
		\caption{\label{fig:App2} Electron and silver ion densities in the $ x $-$ z $ and $ x $-$ y $ slices of the 3D simulation with the shutter and linear polarization (3D W/1xSh - LP).}
	\end{center}
\end{figure}

\setcounter{section}{1}
\label{Appenix1}
As the silver target is ultra-thin, several acceleration mechanisms (and their combination) are acting together during the laser target interaction. From the time evolution of maximal ion energy in figure \ref{fig:App1}(a) at least four different stages can be inferred. The time instants when the stages switch are highlighted by the grid in figure \ref{fig:App1} (times 30 $ T $, 33 $ T $ and 40 $ T $). The electron and silver ion density are shown in the respective time instants (and at time 36 $ T $) in figure \ref{fig:App2}. The density of silver ions was multiplied by $ Z = 40 $ in order to visualize them with the same logarithmic color scale as electrons. The simulation data were recorded with time interval corresponding to laser period $ T $.

The first stage is dominated by the RPA mechanism and last till 30 $ T $. Around this time the laser pulse starts to penetrate through the target as can be seen comparing figure \ref{fig:App1}(b) and \ref{fig:App1}(c) and the temporal evolution of maximal energy stops having an exponential rise, connected to (in principle  unlimited \cite{Bulanov2010Unlim}) RPA and continue with slower rise afterwards. The shell structure (typical for RPA) is developed in the silver ion density distributions in figures \ref{fig:App2}(e) and \ref{fig:App2}(m). The RPA still significantly contributes to ion acceleration in the second stage till time 33 $ T $. This can be seen from figures 2(f) and 2(n) as ions still keep a compact shell structure. 
The ions with highest energy (located at the front of the ion cloud) follow the front electron layer around the position $ x = 7\ \mathrm{\mu m} $. However, the density of the electron layer is decreasing, resulting in slower rise of maximal energy in figure \ref{fig:App1}(a). The electron distribution now consists of high density electron bunches and low density regions around them (figure \ref{fig:App2}(b) and \ref{fig:App2}(j)), thus being partially (relativistically) transparent. Therefore, other mechanisms involving the relativistic transparency, like hybrid RPA-TNSA \cite{Higginson2018_100MeV} and directed coulomb explosion \cite{Bulanov_SS_DCE} also take place at this stage.

 In the third stage between time instants 33 $ T $ and 40 $ T $ the laser pulse propagates through the plasma as can be observed in figure \ref{fig:App2}(c) and \ref{fig:App2}(k). The influence of RPA thus drops significantly and regimes operating with relativistic transparency dominate in this stage. The rise of maximal energy in figure \ref{fig:App1}(a) is noticeably lower in this stage. At time 40 $ T $ most of the laser pulse already left the area where the plasma is located as can be seen in figure \ref{fig:BxSMari}(b). The structures in electron density (figure \ref{fig:App2}(d)) and \ref{fig:App2}(l) expand and slowly disappear. Only a small rise of ion energy is observed afterwards in the last stage.
\section*{References}

\bibliographystyle{iopart-num}
\bibliography{referencesShutter}

\end{document}